\documentclass[preprint2]{aastex}
\usepackage{graphics}
\usepackage{epsfig}
\usepackage{longtable}
\def\gsim{\lower 2pt \hbox{$\, \buildrel {\scriptstyle >}\over
{\scriptstyle \sim}\,$}}
\def\lsim{\lower 2pt \hbox{$\, \buildrel {\scriptstyle <}\over
{\scriptstyle \sim}\,$}}

\newcommand{\xmm}{{\em XMM-Newton}}

\def\chandra{{\em Chandra}}
\def\Chandra{{\em Chandra}}

\def\HI{H{\small I}\ }

\newcommand{\as}{$^{\prime\prime}~$}

\def\cor{\widehat=}

\def\xs{Sombrero}
\shortauthors{}
\shorttitle{}

\begin{document}

\title{\chandra\ and \xmm\ Detection of Large-scale Diffuse X-ray Emission 
from the Sombrero Galaxy}
\author{Zhiyuan Li\altaffilmark{1}, Q. Daniel Wang\altaffilmark{1}, \& Salman Hameed\altaffilmark{2}} 
\altaffiltext{1}{Department of Astronomy, University of Massachusetts, 
710 North Pleasant Street, Amherst, MA 01003}
\altaffiltext{2}{Five College Astronomy Department, Hampshire College, Amherst, MA 01003, USA}
\affil{Email: zyli@nova.astro.umass.edu, wqd@astro.umass.edu, shameed@hampshire.edu}
\begin{abstract}
We  present an X-ray study of the massive edge-on Sa galaxy,
Sombrero (M~104; NGC~4594), based on {\em XMM-Newton} and \chandra\ observations.  
A list of 62 \xmm\ and 175 \chandra\  
discrete X-ray sources is provided, the majority of which
are associated with the galaxy. Spectral analysis is carried out
for relatively bright individual sources and for an accumulated source spectrum. 
At energies $\gtrsim 2$ keV, the source-subtracted X-ray 
emission is distributed similarly as the stellar K-band light and 
is primarily due to the residual emission from discrete sources. 
At lower energies, however, a substantial fraction of
the source-subtracted emission arises from diffuse hot gas
extending to $\sim 20$ kpc from the galactic center.
The galactic disk shows little X-ray emission and instead 
shadows part of the X-ray radiation from the bulge.
The observed diffuse X-ray emission from the galaxy shows a steep spectrum
that can be characterized by an optically-thin thermal plasma with 
temperatures of $\sim$ 0.6-0.7 keV, varying little with radius. 
The diffuse emission has a total luminosity of 
$\sim 3\times 10^{39} {\rm~erg~s^{-1}}$ in the 0.2-2 keV energy range.
This luminosity is significantly smaller than the prediction 
by current numerical simulations
for galaxies as massive as \xs. However, such simulations do not
include the effect of quienscent stellar feedback (e.g., ejecta from evolving
stars and Type Ia supernovae) against the accretion
from intergalactic medium. We argue that the stellar feedback likely plays 
an essential role in regulating the physical properties of hot gas. Indeed, 
the observed diffuse X-ray luminosity of
\xs\ accounts for at most a few percent
of the expected mechanical energy input from Type Ia supernovae. 
The inferred gas mass and metal content are also substantially 
less than those expected from stellar ejecta.
We speculate that a galactic bulge wind, powered primarily by Type Ia supernovae,
has removed much of the ``missing'' energy and metal-enriched gas
from the region revealed by the X-ray observations.
\end{abstract}

\keywords{galaxies: general --- galaxies: individual (Sombrero,
NGC~4594) -- galaxies: spiral --- X-rays: general}

\section{Introduction}

Galactic bulges are an important component of early-type spiral 
galaxies. X-ray studies of the high-energy 
phenomena and processes in galactic bulges 
provide a vital insight into our understanding of 
galaxy formation and evolution. Several facts make 
\xs\ (Table \ref{tab:M104}) an ideal target for such a study: 1) This nearby Sa galaxy is 
massive (circular rotation speed of $\sim 370 {\rm~km~s^{-1}}$)
and bulge-dominated, 
and hence a potential site for probing a large amount of hot gas from
intergalactic accretion (e.g., Toft et al.~2002) and/or  
internal stellar feedback (e.g., Sato \& Tawara 1999);
2) The high inclination of 
the galaxy (84$^\circ$) allows for a clean separation between 
the disk and bulge/halo components; 3) A well-determined 
distance ($8.9 \pm 0.6$~Mpc) of the galaxy minimizes the uncertainty in the
measurement of X-ray luminosities; 
4) As indicated by its very 
low specific far-infrared and diffuse radio fluxes (Bajaja et al. 1988), 
the galaxy shows little indication for recent star formation, 
minimizing the possibility of heating and/or gas ejection 
from the galactic disk; 5) The galaxy is isolated
and thus uncertainties resulting from galaxy interaction are miminal. 
Therefore, \xs\ is particularly well-suited for
an X-ray study of high-energy stellar and interstellar products in a 
galactic bulge and their relationship to the galactic disk and to the 
intergalactic environment.

Existing X-ray studies of \xs\ have focused on its discrete X-ray sources. 
Di Stefano et al.~(2003) reported the detection of 122 
X-ray sources, based on a \Chandra\ ACIS-S observation of the 
galaxy. In particular, they classified a population of very soft X-ray 
sources, which tend to concentrate in the core region of
the galactic bulge. Wang (2004) conducted a careful analysis of 
the luminosity function of the discrete X-ray sources detected
from the same observation by correcting for
incompleteness and Eddington bias in the source detection and by removing statistical interlopers in the field. The X-ray behavior of the central AGN has been studied by Pellegrini et al.~(2003), based on an \xmm\ observation as well as the {\sl Chandra} data. 

\begin{figure}[!htb]
 \vskip -1cm 
 \centerline{
      \epsfig{figure=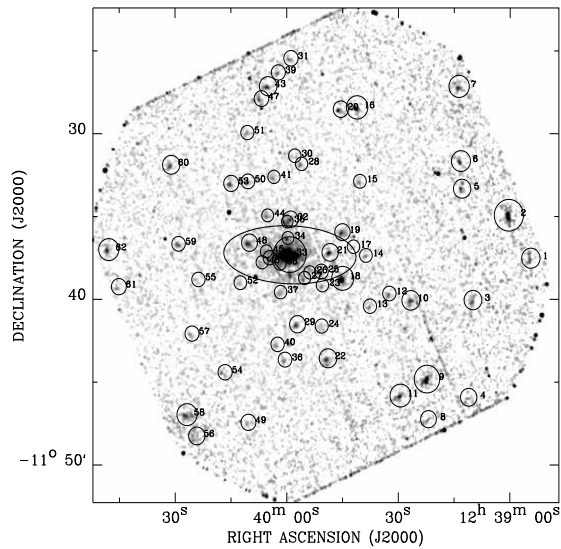,width=0.5\textwidth,angle=0}
    }
  \caption{EPIC-PN intensity image of the M~104 field in the 0.5-7.5 keV band after 
a flat-fielding. An adaptively
smoothed background has been subtracted from
the image to highlight discrete sources which are outlined 
with circles for source-removal (see \S~\ref{subsec:spat_anal}).
The source numbers (Table \ref{tab:pn_source_list}) are also marked. 
The ellipse ($8\farcm7{\times}3\farcm5$) illustrates 
the optical $I_B=25 {\rm~mag~arcsec^{-2}}$ isophote of the galaxy.
\label{fig:sou_pn}}
\end{figure}

\begin{figure*} 
\centerline{
\psfig{figure=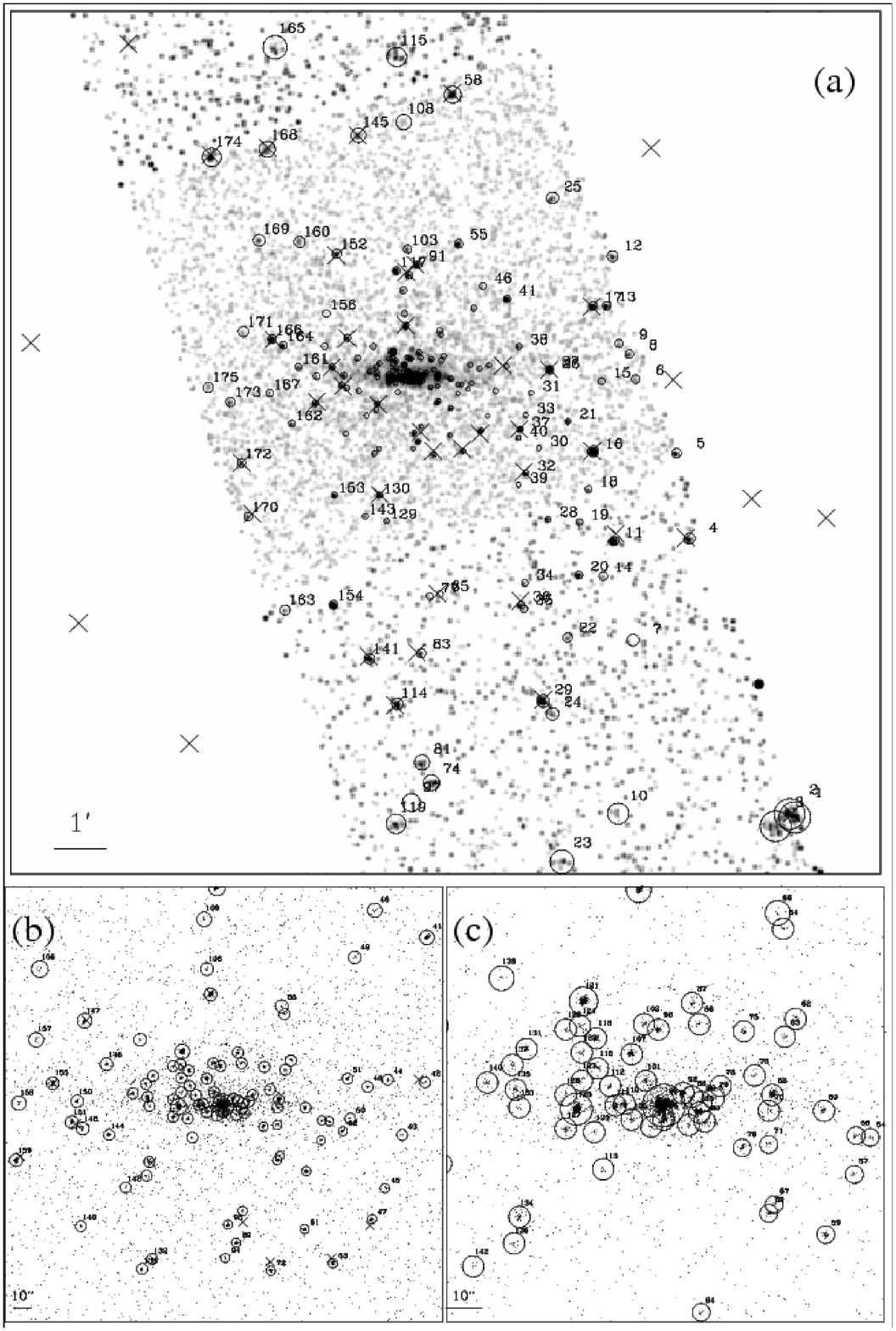,height=7.5in,angle=0, clip=}
}
\caption{ACIS-S 0.3-7 keV band intensity images: 
(a) the whole source detection field, over which the image is
smoothed with a Gaussian of FWHM equal to 3\farcs9; (b) 
the inner $\sim 4^\prime{\times}4^\prime$ region around the center of M~104; (c) the very central $\sim 2^\prime{\times}2^\prime$ region around the galactic center. 
Detected X-ray sources (Table \ref{tab:acis_source_list}) are outlined
with circles for source-removal (see \S~\ref{subsec:spat_anal}). Positions of sources detected by
the EPIC-PN are marked with {\sl crosses}.
}
 \label{fig:sou_acis}
\end{figure*}

\begin{figure}[!htb]
  \centerline{
      \epsfig{figure=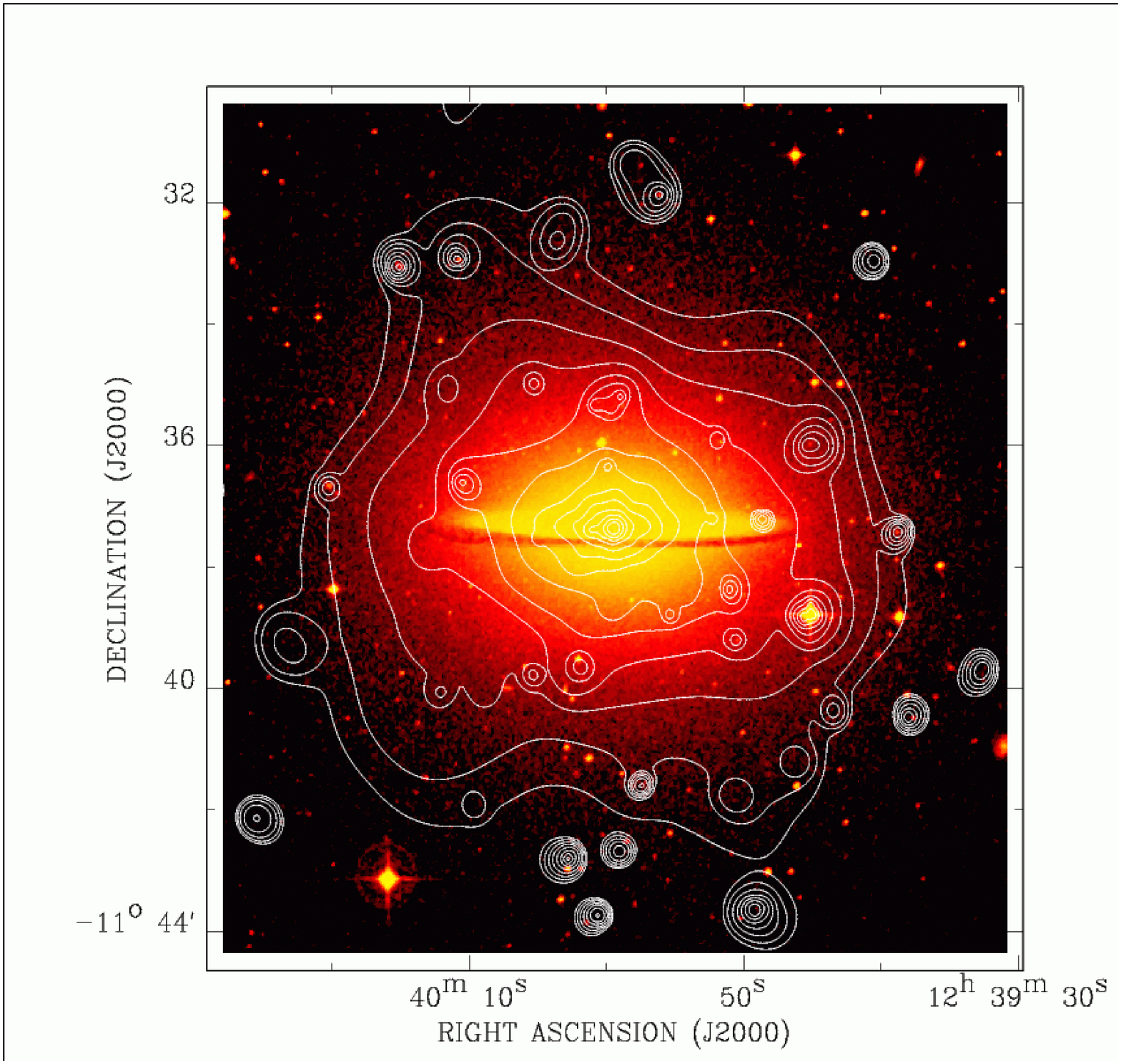,width=0.5\textwidth,angle=0}
    }
  \caption{EPIC-PN 0.5-2 keV intensity contours overlaid on the 
digitized sky-survey blue image of M~104. The X-ray intensity 
is adaptively smoothed with the CIAO {\sl csmooth} routine with 
a signal-to-noise ratio of $\sim3$. The contours are at 
(1.4, 1.8,   2.6,  4.2,    7.4,   13.8, 
      27, 52, 103, 206, 411,  820, and
      1640)$\times 10^{-3} {\rm~cts~s^{-1}~arcmin^{-2}}$ above a local
background level of 2.0$\times 10^{-3} {\rm~cts~s^{-1}~arcmin^{-2}}$.
    \label{fig:x_o_c}}
\end{figure}

We here report a systematic analysis of the 
\xmm\ and \chandra\ observations (Figs.~\ref{fig:sou_pn} and \ref{fig:sou_acis}),
focusing on the study of diffuse X-ray emission in \xs.   
The \chandra\ data, with superb spatial resolution, are
well-suited for the study of the galaxy's inner region
where the X-ray source density is high. However, the field of view (FoV) of
the \chandra\ ACIS-S, especially that of the S3 chip ($\sim 
8^\prime \times 8^\prime$), does not provide a full coverage of the large-scale
X-ray emission of the galaxy
(cf. Fig.~\ref{fig:x_o_c}). The \xmm\ EPIC observation, on the other
hand, has a substantially larger FoV, allowing us to probe the 
extent of the global diffuse X-ray emission.
The combination of the two observations thus provides us with the
most comprehensive X-ray view of the galaxy. 

\section{Observations and Data Reduction} {\label{sec:data}}
Our data calibration procedures have been
detailed in previous works which dealt with similar \chandra\ and \xmm\ 
observations (e.g., Wang et al. 2003; Li et al.~2006). 
Here we summarize the essential aspects that are specific to
the current data.

\subsection{\chandra\ observations}
The \chandra~ACIS-S observation of \xs\ (Obs.~ID.~1586) was taken  
on May 31, 2001, with an exposure of 18.8 ks. 
Our work uses the data primarily from the on-axis S3 chip, although part of
the adjacent FI chips (S2 and S4) are also included in the
source detection.
We reprocessed the \chandra\ data, using CIAO, version 3.2.1 and the latest calibration files. 
We also removed time intervals with significant background flares, i.e.,
those with count rates $\gtrsim 3\sigma$ and/or a factor of 
$\gtrsim 1.2$ off the mean background level of the observation.
This cleaning resulted in an effective
exposure of 16.4 ks for subsequent analysis. We created 
count and exposure maps in the 0.3-0.7, 0.7-1.5, 1.5-3, and 3-7 keV bands.
Corresponding background maps were created from the ``stowed background'' data, which 
contain only events induced by the instrumental background.
A normalization factor of $\sim1.05$ was applied to the exposure of this ``stowed background''
data in order to match its 10-12 keV count rate with that of Obs.~1586. 

\subsection{\xmm\ observations}
The \xmm\ EPIC observation of \xs\ (Obs.~ID 0084030101) 
was taken on December 28, 2001, with the thin filter and with a total exposure of 43 ks.
We calibrated the data using SAS, version 6.1.0, together with
the latest calibration files. In this work, 
we only use the EPIC-PN data. 
We found that a large fraction of the observation was 
strongly contaminated by cosmic-ray-induced flares. 
To exclude these flares, we
removed time intervals with count rates greater than 
11 ${\rm cts~s^{-1}}$ in the 0.2-15 keV band, about a factor of 1.2 above the quiescent 
background level, and some additional intervals with residual flares found in sub-bands.
The remaining exposure is only 10.8 ks for the PN.  We then 
constructed count and exposure maps in the 0.5-1, 1-2, 2-4.5, 
and 4.5-7.5 keV bands for flat-fielding. We also created corresponding 
background maps from the ``filter wheel closed'' (FWC) data
, chiefly for instrumental X-ray background subtraction.
However, we found that at energies above 5 keV the spectral shape of the 
instrumental background of Obs.~0084030101 is apparently different
from that of the FWC data, making a simple normalization inapplicable.   
Therefore, the FWC data are only used in producing large-scale images.
Background adoption for spectral analysis will be further discussed in \S~\ref{subsec:spec_anal}.

\section{Discrete X-ray sources} \label{sec:ps}
Fig.~\ref{fig:x_o_c} shows the overall 0.5-2 keV X-ray intensity image
of \xs\ obtained from the PN. The morphology appears more-or-less 
symmetric, reminiscent of the optical light distribution of the galaxy.
The X-ray emission likely represents a combined contribution from discrete 
sources and truly diffuse hot gas. We first detect individual sources
and characterize their properties. Then we try to isolate and study the
diffuse X-ray component in \S~\ref{sec:diffuse}.  

\subsection{Source dection and astrometry correction} {\label{subsec:detect}}
We detect 175 \chandra\ and 62 \xmm\ discrete X-ray sources.
Tables~\ref{tab:acis_source_list} and \ref{tab:pn_source_list} summarize the detection results. 
The source detection is carried out for each observation in 
the broad (B), soft (S), and hard (H) bands, defined differently for
the ACIS-S and PN data, as noted
in the tables. Following the procedure detailed in Wang (2004), we
use a combination of source detection 
algorithms: wavelet, sliding-box, and maximum likelihood 
centroid fitting. The map detection and 
the maximum likelihood analysis are based on data within the 50\% PSF
energy-encircled radius (EER) for the PN and the 90\% EER for the 
ACIS-S. The accepted sources all have a local
false detection probability $P\le 10^{-6}$.  
For ease of reference, we will 
refer to X-ray sources detected in the
PN and the ACIS-S with prefixes XP and XA,
respectively (e.g., XP-13). 

Although the pointing uncertainty of \chandra\ is {\sl on average} less
than $\sim 1$\as, it is still desirable to quantify and possibly improve 
the astrometric accuracy of any particular observation.
We first use the Two Micron All Sky Survey (2MASS) 
All-Sky Catalog of Point Sources (Cutri et al.~2003) to find potential 
near-IR counterparts for an astrometric calibration.  The astrometry of the 2MASS 
objects is generally much better ($\sim 0\farcs1$). 
We cross-correlate the spatial positions of the objects in the catalog 
with those of the X-ray sources listed in Tables~\ref{tab:acis_source_list} and \ref{tab:pn_source_list}.
For each ACIS-S source, we use a matching radius of twice its position uncertainty, with lower and upper limits of 1\as and 2\as. The lower limit is set
to account for any systematic errors in the X-ray
source positions, whereas the upper limit is to minimize the
probability of chance coincidence. Similar calibration is also done 
for the PN sources outside the ACIS-S FoV
(Fig.~\ref{fig:sou_acis}), with lower and upper matching radii of 2\as and 4\as.

The calibration gives five {\sl Chandra}/2MASS position coincidences.
We estimate the required astrometric correction to be 0\farcs4 to the west and
0\farcs5 to the north, based on a $\chi^2$
fit to the R.A. and DEC. offsets of the five matched {\sl Chandra}/2MASS pairs. The
correction is insensitive (with changes $\lesssim 0\farcs1$) to the 
exclusion of any one of the entries in the fit. 
The correction improves the $\chi^2/d.o.f.$ from 21/10 to 4/8.  
After correcting for the X-ray astrometry, one additional position coincidence (XA-149) is found.

Table~\ref{tab:sou_id} presents
the matching results, including the position offset of each match, together
with the expected position uncertainties quoted from Tables 1 and 2; 
there is no match with multiple 2MASS objects.  
The table also includes 
the J, H, and K$_s$ magnitudes of the matched 2MASS objects; the 
$3\sigma$ limiting sensitivities of the catalog are 17.1, 16.4 and 15.3 
mag in the three bands. We estimate the expected 
number of chance projections of 2MASS objects within the 
matching regions to be $\sim 0.2$ for
the ACIS-S sources and 0.5 for the PN sources, based on 
the surface number density of 2MASS objects within annuli
of 4\as-15\as radii around the X-ray sources. Therefore, it is possible 
that a couple of the matches may just be such chance projections.

The source locations are marked in Figs.~\ref{fig:sou_pn} and \ref{fig:sou_acis}. 
Essentially all PN sources within
the field of Fig.~\ref{fig:sou_acis}a are also detected in the ACIS-S data.
All relatively bright ACIS-S sources, except for those in the nuclear region, 
are detected in the PN data. These consistencies indicate no strong 
variability of the sources between the two observations. 
Source confusion is serious for the PN data, 
because of the limited spatial resolution. Some of the
PN detections represent combinations of multiple discrete sources,
e.g., XP-42 is a combination of XA-148 and 151. This is particular the case
in the nuclear region (Fig.~\ref{fig:sou_acis}c). 

We note that the source detection limit is significantly higher in the 
PN data than in the ACIS-S data. The majority of detected sources are of
two populations: sources associated with the galaxy and extragalactic sources mostly being background AGNs.
Applying the luminosity function (LF) of the AGNs obtained by Moretti et al.~(2003),
we estimate the number of detected AGNs to be 15.5 (2.3) in the ACIS-S (PN) FoV.

\subsection{Individual spectra of bright sources} 
\begin{figure*}[!htb]
  \centerline{\epsfig{figure=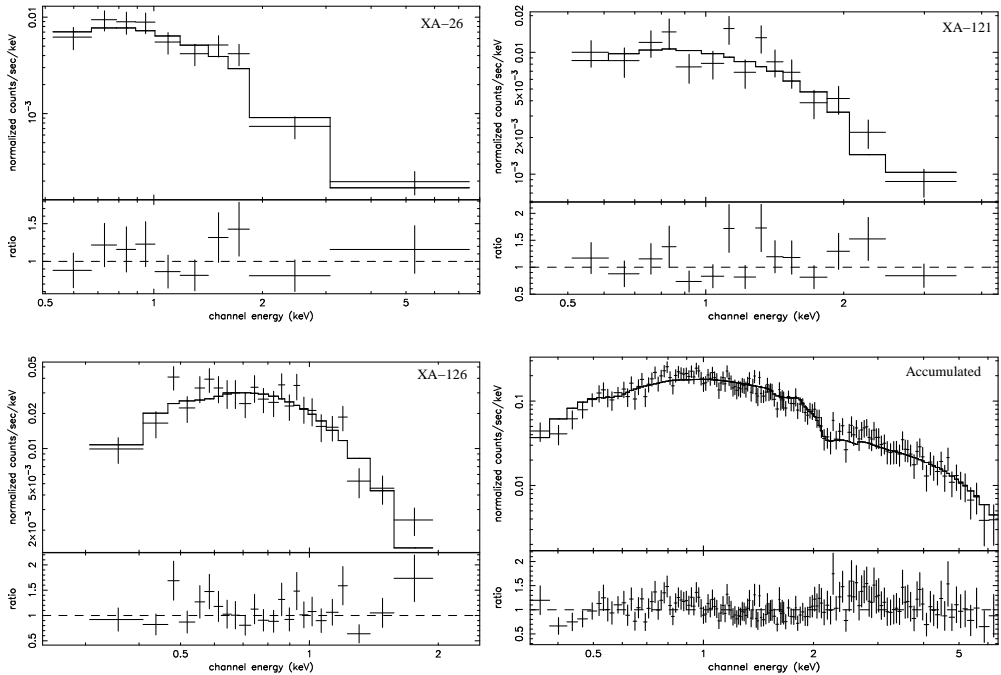,width=0.8\textwidth,angle=0}}
  \caption{ACIS-S spectra of discrete sources in the field of M~104 
: (a)-(c) individual spectra of three bright sources; (d) an accumulated spectrum
of sources within the D$_{25}$ ellipse. Best-fit models to the spectra are also 
shown. The lower panels show the data-to-model ratios.
\label{fig:sou_spec}}
\end{figure*}

{\label{subsec:spec_ps_br}}
There are eight bright sources in the ACIS-S field as deteceted with a count rate
higher than 0.01 ${\rm~cts~s^{-1}}$. These are XA-10, 15, 26, 28, 96, 98, 121 and 124
(Table~\ref{tab:acis_source_list}). Among them, XA-15 is a foreground star; XA-10 and XA-28
are located outside the S3 chip;
X-96 is the nucleus which have been studied by Pellegrini et al.~(2003);
X-98 is located within 3$^{\prime\prime}$ from the nucleus. 
We perform spectral fit to individual ACIS-S spectra of the rest three sources,
each extracted from a circular region
of twice the 90\% EER.
Corresponding background spectra are extracted from the source-free vicinity of each source.   
The spectra of XA-26 and XA-121 are fitted by an absorbed power-law, whereas
the spctrum of XA-124, being very soft, is fitted by an absorbed blackbody.
The fit results, summerized in Table \ref{tab:sou_spec}, are statistically
consistent with those obtained by Di Stefano et al. (2003).
The spectra and the best-fit models are shown in Fig.~\ref{fig:sou_spec}a-\ref{fig:sou_spec}c.

\subsection{Accumulated source spectrum} {\label{subsec:spec_ps}}
We further obtain an accumulated ACIS-S spectrum of the sources to characterize their
average spectral property (Fig.~\ref{fig:sou_spec}d). The spectrum 
is extracted from sources within the $I_B=25 {\rm~mag~arcsec^{-2}}$ isophote 
($D_{25}$ ellipse; 8\farcm7$\times$3\farcm5), except for the nuclear source and the 
three bright sources discussed above. The total number of included sources is $\sim$110. 
For each source, a circular region of twice the 90\% EER is adopted for accumulating the spectrum.
A background spectrum is extracted from the rest region of the ellipse.
In the PN data, only ten sources are detected within the $D_{25}$ ellipse and 
four of them are located within 1\farcm5 from the galactic center, where the
emission of the nucleus largely affects. 
Therefore, we do not analyze an accumulated source spectrum from the PN.

We use an absorbed power-law model to fit the accumulated spectrum, with
the absorption being at least that supplied by the Galactic foreground. 
The model offers an acceptable fit to the spectrum (Table~\ref{tab:sou_spec}), giving a best-fit photon index 
of $1.51^{+0.10}_{-0.09}$ and a 0.3-7 keV intrinsic luminosity of $\sim2.6{\times}10^{40}{\rm~ergs~s^{-1}}$. 
All quoted errors in this paper are at the 90\% confidence level.
The slope of the power-law is typical for composite X-ray spectra of low-mass X-ray binaries observed in nearby galaxies 
(LMXBs; e.g., Irwin, Athey \& Bregman 2003). We note that none of the included sources 
contributes more than 5\% of the total counts to the accummulated spectrum. Therefore, the spectrum, along with the fitting
model, can be used to characterize the average spectral property of sources.

\section{The source-subtracted X-ray emission}  {\label{sec:diffuse}}
Our main interest here is in the diffuse X-ray emission from \xs. 
A first step towards isolating the diffuse emission is to subtract the 
detected discrete sources from the images.
To do so, we exclude regions enclosing twice the 50\% (90\%) EER around each PN (ACIS-S) source
with a count rate ($CR$) $\lesssim 0.01 {\rm~cts~s^{-1}}$.
For brighter sources, 
a factor of $1+{\rm log}(CR/0.01)$ is further multiplied to the source-subtraction radius. 
Our choice of the regions is a compromise between excluding a bulk of the source
contribution and preserving a sufficient field for the study of 
the source-subtracted emission.
With the above criteria about $80\%$ ($95\%$) of photons from individual
sources are removed from the PN (ACIS-S) image. 

The source-subtracted emission presumably consists of two components: the emission of 
truly diffuse gas and the collective discrete contributions from
the residual emission of detected sources and the emission 
of undetected sources below our detection limit. In practice, the discrete component
can be constrained from its distinct spatial distribution and spectral property. 
Below we isolate the two components and characterize the properties of the diffuse emission.

\subsection{Spatial properties} {\label{subsec:spat_anal}}
\subsubsection{Surface intensity profiles} 
\begin{figure*}[!htb]
  \centerline{
   \epsfig{figure=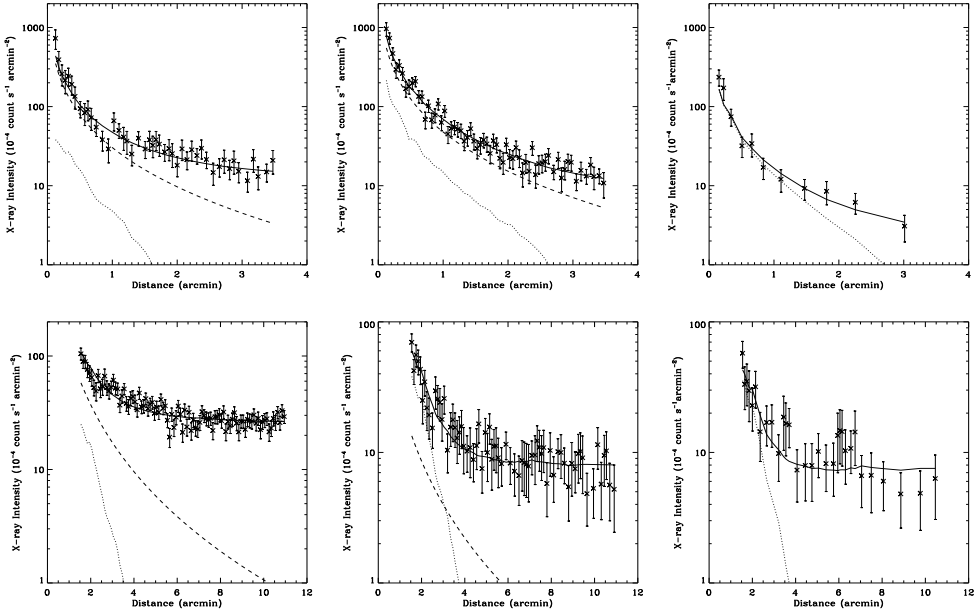,width=0.8\textwidth,angle=0}
  }
\caption{Radial surface intensity profiles of the instrumental 
background- and detected source-subtracted emission from 
M~104. {\sl Top}: ACIS-S profiles in the 0.3-0.7 keV ({\sl left}), 0.7-1.5 keV ({\sl middle}) and 1.5-7 keV ({\sl right}) bands.
{\sl Bottom}: PN profiles in the 0.5-1 keV ({\sl left}), 1-2 keV ({\sl middle}) and 2-7.5 ({\sl right}) bands.
The solid curves present model characterizations: a normalized  
K-band radial profile for emission from discrete sources (dotted
curves), a de Vaucouleur's law (dashed curves) for emission from diffuse hot gas, 
and a local constant cosmic background. See text for details.
}
\label{fig:rbp}
\end{figure*}

{\label{subsubsec:rbp}}
We construct instrumental background-subtracted and exposure-corrected 
galactocentric radial surface intensity profiles for the
source-subtracted emission, in the soft (0.5-1 keV for the PN; 0.3-0.7 keV for 
the ACIS-S), 
intermediate (1-2 keV for the PN; 0.7-1.5 keV for the ACIS-S)
and hard (2-7.5 keV for the PN; 1.5-7.0 keV for the ACIS-S) bands (Fig.~\ref{fig:rbp}). 
While the ACIS-S instrumental background is determined from
the ``stowed background'' data, the instrumental background rates in the PN bands
are predicted from the spectral fit to a local PN background spectrum (see \S~\ref{subsec:spec_anal}).
Spatial binning of annuli is adaptively adjusted to achieve a
signal-to-noise ratio better than 3, with a minimum step size of
6$^{\prime\prime}$ for the PN and 3$^{\prime\prime}$ for the ACIS-S.
For the PN profiles, the central 1\farcm5 is heavily
contaminated by the emission from the nucleus. Thus our 
analysis for the PN data is restricted to radii beyond 1\farcm5.
The ACIS-S data, while being capable to probe the central region, 
are limited by its FoV. Therefore we restrict our analysis of the ACIS-S
profiles within 3\farcm5, a maximal radius where complete annuli can be extracted.

It is known that \xs\ has a prominent dust lane (e.g., Knapen et al.~1991; cf.~Fig.~\ref{fig:x_o_c}),
which may significantly absorb soft X-rays from the galaxy and
hence introduce a bias to the bin-averaged intensity.
Therefore when constructing the intensity profiles we exclude a region 
encompassing the dust lane. We use the digitized sky-survey blue
image of the galaxy to map such a region of extinction, in which a pixel is adopted as the region
bounary if it is dimmed by a factor $\geq$1.5 compared to the adjacent bright pixel.
Visual inspection on the ACIS-S image indicates that 
our adopted region is coincident with a region of few registered soft X-ray photons.

To constrain the discrete component,
we assume that its spatial distribution
follows the near-IR light of the galaxy,
which can be determined from the 2MASS K-band map (Jarrett et al.~2003).
We first exclude from the map bright foreground stars and circular regions used
for subtracting the discrete X-ray sources.
The K-band radial intensity profile is then produced in the same
manner as for the X-ray profiles. It is reasonable
to assume that the discrete component dominates
the X-ray emission in the hard band. Thus we use the K-band profile 
to fit the X-ray hard band profiles, constructed from the PN and ACIS-S data.
The fitting parameters are 
the X-ray-to-K-band intensity ratio of the underlying
stellar content ($I_s$) and a constant intensity ($I_b$)
to account for the local cosmic X-ray background. 
We find that the X-ray
hard band profiles can be well characterized by the K-band profile (Fig.~\ref{fig:rbp}; Table~\ref{tab:rbp}).
 
We further assume that the discrete component
has a collective spectral property same as that modeled for 
the detected sources (\S~\ref{subsec:spec_ps}). 
This allows us to use the hard band intensity to constrain
the discrete component in the soft and intermediate bands.
The diffuse component is then determined for
these two bands by subtracting the discrete component
from the total intensity profile.
We then fit the radial distribution of the diffuse component with a de Vaucouleur's law:
\begin{equation}
I(R) = I_g~e^{-7.67(R/r_e)^{1/4}},
\label{eq:deVau}
\end{equation}
where $R$ is the projected galactocentric radius, $r_e$ the half-light radius
and $I_g$ the central surface intensity.
A paramter $I_b$ is also included to account for the local cosmic X-ray background. 
Due to the partial coverage of the overall distribution by each profile,
we require in the fit that the half-light radii be identical for all profiles.
We find that this 
characterization offers good fits to the profiles (Fig.~\ref{fig:rbp}; Table~\ref{tab:rbp}). 
The best-fit half-light radius is 2\farcm6$^{+1\farcm4}_{-0\farcm9}$.
In comparison, the K-band half-light radius is 
$\sim$ 1$^\prime$ ($\sim$ 2.6 kpc; Jarrett et al.~2003).
This suggests that the distribution of hot gas is substantially more extented than that of the stellar content. 

\begin{figure}[!hbt]
\centerline{ 
\psfig{figure=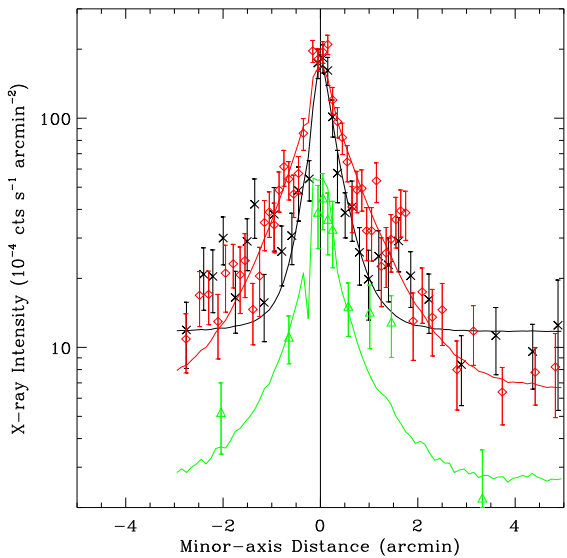,width=0.5\textwidth,angle=0, clip=}
}
\caption{ACIS-S intensity distribution along the direction
perpendicular to the disk of M~104, in the 
0.3-0.7 keV ({\sl black crosses}), 0.7-1.5 keV ({\sl red diamonds}), 
and 1.5-7 keV ({\sl green triangles}) bands.
The full width along the direction parallel to the major axis 
used for averaging the intensity is 4$^\prime$ ($\sim$10 kpc). 
The adaptive steps along
the minor axis achieve
a signal-to-noise ratio greater than 3, with
a minimum of stepsize of 6$^{\prime\prime}$. The vertical
line represents the position of the major 
axis of the disk, whereas the horizontal axis 
marks the vertical distance along the minor axis
(south as negative).
}
 \label{fig:rvp}
\end{figure}

We also construct vertical intensity profiles of the source-subtracted emission  
along the galaxy's minor axis for the ACIS-S 0.3-0.7, 0.7-1.5 and 1.5-7 keV 
bands (Fig.\ref{fig:rvp}). 
In general, the intensity decreases 
rapidly with the off-disk distance. 
We follow the above procedure to decompose the diffuse and discrete components
of the vertical profiles. An expotential law, i.e., $I(z) = I_g~e^{-|z|/z_0}$,
is used to fit to the vertical distribution of the diffuse component.
The scale height $z_0$ is allowed to be different between the south and north
sides of the midplane.
The fit is marginally accetable, with excess existing at $\sim2^{\prime}$ 
from the midplane on both sides.
Fit results (Table~\ref{tab:vbp}) show that in each band
there is no significant asymmetry in the intensity distribution with respect to 
the midplane.
The best-fit scale height in the soft band ($\sim$ 1/4 arcmin) is less than that
in the intermediate band ($\sim$ 1/3 arcmin), indicating that emission is softer in the central region than in the extraplanar region.
When the above fit is restricted to a vertical distance $\ge$ 0\farcm5, the best-fit scale
heights for the soft and intermediate bands are nearly identical ($\sim 1^\prime$).
This is evident that the temperature of hot gas around the disk plane is lower than that in the bulge.

\subsubsection{Inner region and substructures} 
\begin{figure}[!htb]
\centerline{
\psfig{figure=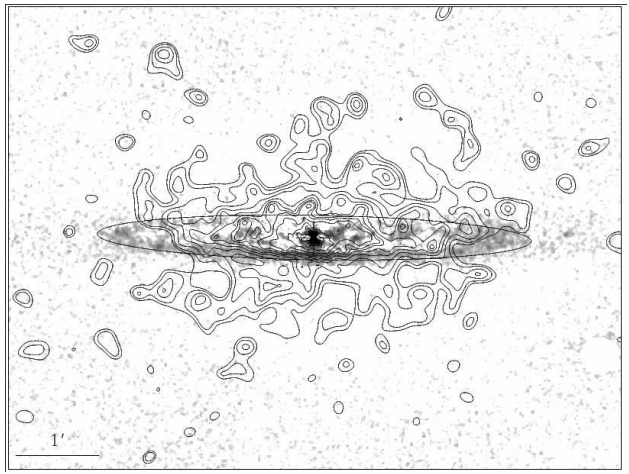,width=0.5\textwidth,angle=0, clip=}
}
\caption{ACIS-S 0.3-1.5 keV intensity contours overlaid on the 
continuum-subtracted H$\alpha$ image of M~104. The X-ray intensity
is smoothed adaptively with a count-to-noise ratio of 4 after source-subtraction. 
}
\label{fig:ha_x}
\end{figure}

{\label{subsubsec:center}}
We use the ACIS data to probe the diffuse X-ray properties in the 
inner region of the galaxy. 
We fill the holes from the 
source removal with the values interpolated from surrounding bins. 
Fig.~\ref{fig:ha_x} shows ``diffuse'' X-ray intensity contours,
which are substantially less smoothed than presented in Fig.~\ref{fig:x_o_c}. 
There are considerable substructures in the inner region.
Inner contours are extended more to the
north than to the south (where strong intensity gradients are found), 
indicating a heavier absorption of X-ray emission 
to the south. This is clearly due to the prominent dust 
lane that lies at the 10{\as}-25{\as} range to the south 
of the major axis (Knapen et al. 1991). The
intensity contours also become strongly elongated along the galactic disk. 

Fig.~\ref{fig:ha_x} also presents in grey scale 
a continuum-subtracted H$\alpha$ image of \xs, 
obtained with the 0.9 meter telescope at Kitt Peak National Observatory 
in 1999. The details of observations are presented elsewhere
(Hameed \& Devereux 2005). H$\alpha$ emission is distributed, 
primarily, in an annulus, and individual 
HII regions can be identified on the ring. There is some 
H$\alpha$ emission within the ring, but it is difficult 
to tell from the image if the emission is diffuse or if it contains HII regions. 
The H$\alpha$ ring follows the optical dust lane but is 
located on its inner side. High extinction 
possibly obscures ionized emission from the dust lane itself. 

Total H$\alpha$ flux for \xs, uncorrected for 
internal or external extinction, is calculated to be 
$\sim 0.8 \times 10^{-12}{\rm~ergs~s^{-1}~cm^{-2}}$, which translates 
to a luminosity of $\sim 7.6 \times$ 10$^{39}{\rm~ergs~s^{-1}}$. 
Using Kennicutt's (1998) formula, we derive a star formation rate of 
$\sim 0.1 {\rm~M_{\odot}~yr^{-1}}$, which is lower than the 
average star formation rate (0.9 ${\rm~M_{\odot}~yr^{-1}}$) 
for early-type spirals (Hameed \& Devereux 2005).  

Fig.~\ref{fig:ha_x} shows that X-ray intensity drops abruptly in the field 
covered by the front side of the H$\alpha$ disk, corresponding to the
inner region of the cold gas disk of the galaxy. 
This means that the H$\alpha$-emitting
region does not contribute appreciable amounts to X-ray radiation.  
In contrast, good H$\alpha$/X-ray correlation is typically seen in 
late-type spirals (e.g., Strickland et al. 2004; Wang et al. 2003). In fact, 
the diffuse X-ray intensity in \xs\ is so low in the
field covered by the front side of the cool gas disk that the disk must be absorbinga large fraction of X-ray emission from the region beyond.

\begin{figure}[!htb]
\centerline{
\psfig{figure=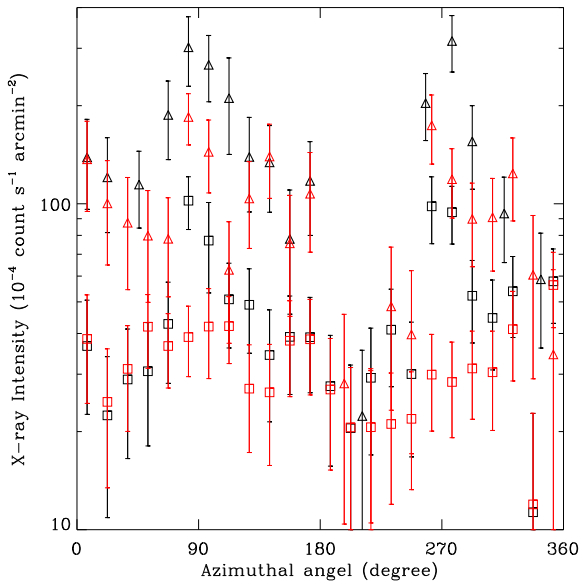,width=0.5\textwidth,angle=0, clip=}
}
\caption{Azimuthal diffuse intensity distributions in the ACIS-S 0.3-1.5 keV band, 
averaged within annuli with inner-to-outer radii of 30$^{\prime\prime}$-1$^\prime$ (black triangles) 
and 1$^\prime$-2$^\prime$ (black squares). As comparison are similar distributions
(red triangles and squares) within elliptical annuli of an axis ratio of 2/3.
The angle is counterclockwise from the minor axis (north).
Contribution from unresolved sources is subtracted according to the underlying K-band 
light (\S~\ref{subsec:spat_anal}; Table~\ref{tab:rbp}).
Adaptive binning is taken to have a minimum step of 15$^\circ$ and to achieve a signal-to-noise
ratio better than 3. 
}
\label{fig:surbaz}
\end{figure}

We probe the azimuthal variation of diffuse emission in the inner region. Fig.~\ref{fig:surbaz} shows
the azimuthal ACIS-S 0.3-1.5 keV intensity distributions.
The distributions deviate from axisymmetry significantly. But the deviations are largely coupled with 
the orientation of the bulge (0$^\circ$ aligns with the minor axis). 
When the azimuthal intensity distributions are measured within elliptical annuli with an axis ratio
similar to that of the bulge (Fig.~\ref{fig:surbaz}), the deviations are significantly reduced, 
with smaller scale fluctuations remaining in certain azimuthal ranges,
especially in the inner region. For example, dips present at $\sim200^\circ-250^\circ$ and $\sim330^\circ-350^\circ$ find 
their counterparts in Fig.~\ref{fig:ha_x}. At larger radii, only moderate deviations 
from axisymmetry can be seen from the azimuthal intensity distributions
for the PN data. When an axis ratio of 0.8 is adopted to reflect the geomoetry of the bulge, most of the deviations vanish 
and no substantial fluctuations are present. This is evidence that the diffuse emission is nearly axisymmetric at large scale. 

\subsection{Spectral properties of the diffuse X-ray emission} {\label{subsec:spec_anal}}

With the above spatial properties in mind,
we perform spectral analysis of source-subtracted emission 
from a series of concentric annuli around the galactic center.
Specifically, spectra are extracted from two annuli with 
inner-to-outer radii of
30$^{\prime\prime}$-1$^\prime$ and 1$^\prime$-2$^\prime$ for the ACIS-S data
and two annuli of 2$^\prime$-4$^\prime$ and 4$^\prime$-6$^\prime$ for the PN data.
The dust lane region (\S~\ref{subsubsec:rbp}) is excluded from the spectral extraction.

Two factors complicate the background determination in our spectral analysis.  
First, the sky location of \xs\ is on the edge of the North Polar Spur (NPS),
a Galactic soft X-ray-emitting feature (Snowden et al. 1995). 
The NPS introduces an enhancement to the local background, 
particularly at low energies.
Secondly, the X-ray emission from \xs\ extends to
at least 6$^\prime$ from the galactic center (\S~\ref{subsubsec:rbp}). 
Thus a local background cannot be extracted for the ACIS-S data. 
While the PN FoV still allows for a local background, 
the vignetting effect at large off-axis angles needs to be properly
corrected for in the background subtraction. Generally, this can be
achieved with the ``double-subtraction'' procedure: a first subtraction
of the non-vignetted instrumental background followed by a second
subtraction of the vignetted local cosmic background. Such a procedure
relies on the assumption that the template instrumental background
can effectively mimic that of a particular observation.

\begin{figure}[h] 
\centerline{
\psfig{figure=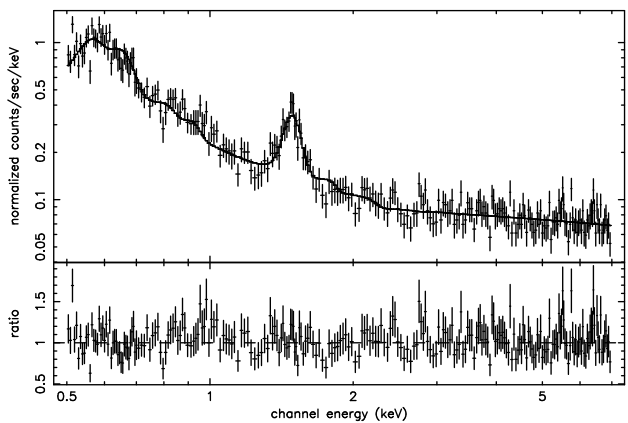,width=0.5\textwidth,angle=0, clip=}
}
\caption{The PN background spectrum with the best-fit model. 
Note for a strong instrumental line at the energy of $\sim$ 1.5 keV. See text for details.}
\label{fig:pn_back}
\end{figure}

We intend to perform the ``double-subtraction'' procedure to determine the background.
First we extract the PN background spectrum from a source-subtracted annulus 
with inner-to-outer galactocentric radii of 8$^\prime$-11$^\prime$,
a region containing little emission from the galaxy (Fig.~\ref{fig:rbp}).
However, at energies $\ge$ 5 keV, where the instrumental background
is predominant, the local background spectrum is found to be significantly harder 
than the spectrum extracted from the FWC data. 
Therefore, we decide to characterize the local background spectrum of PN, both instrumental
and cosmic, by a combination of plausible components.
To model the instrumental background, a broken power-law plus several Gaussian lines
is applied (Nevalainene, Markevitch \& Lumb 2005). 
The modeling of the cosmic background consists of three components. 
Two of them are thermal (the APEC model in XSPEC), representing
the emission from the Galactic halo (temperature $\sim$ 0.1 keV) and the 
NPS (temperature $\sim$ 0.25 keV; Willingale et al.~2003), respectively.
The third component is a power-law with the photon index fixed at 1.4, 
representing the unresolved extragalactic X-ray emission (Moretti et al.~2003).     
Our combined model results in a good fit to the local background spectrum. We note that
the decomposition of the local background is not unique, especially
at lower energies ($\lesssim$ 1 kev).
We verify our modeling by the fact that the fitted parameters of these commonly 
used cosmic components are in good agreement
with independent measurements (e.g., Willingale et al.~2003; Moretti et al.~2003). 
The background spectrum in the 0.5-7 keV range, grouped to have 
a minimum number of 30 counts in each bin, is shown in Fig.\ref{fig:pn_back}.
The model, scaled according to the corresponding sky areas,
is included in the following fit to the PN spectra of source-subtracted emission.

The spectral shape of the ACIS-S instrumental background is rather stable at energies $\ge$
0.5 keV{\footnote{http://cxc.harvard.edu/contrib/maxim/stowed/}. Also,
as the ACIS-S spectra of source-subtracted emission are extracted within the central
 2$^\prime$ where the surface intensity is peaked, a small uncertainty in the instrumental 
bakcground subtraction would only cause a minor effect in the analysis. Therefore,
we directly subtract a ``stowed backgroud'' spectrum from the ACIS-S spectra of 
source-subtracted emission and group them to achieve a signal-to-noise ratio better than 3. 
The remaining cosmic X-ray background are modeled with the same components as
for the PN spectra. We note that an additional factor of 0.43, estimated from the LF obtained
by Moretti et al.~(2003), is multiplied to the scaling of the extragalactic component 
in order to account for the lower source detection limit in the ACIS-S data (Wang 2004).

The spectra show a clear line feature 
at $\sim$0.9 keV (Fig.~\ref{fig:spectra_all}), presumably
due to the Fe L-shell complex contributed by the hot gas, 
while at energies above 1.5 keV
the spectra are dominated by the residual emission of
discrete sources. 
We account for the discrete contribution with a power-law model (PL) with
a fixed photon index of 1.51 (Table~\ref{tab:sou_spec}), 
again assuming that its collective spectral shape 
is same as that of the detected sources.
This PL, combined with a thermal plasma emission model (APEC) characterizing
the emission of hot gas,
is used to simultaneously fit the four spectra.
Both components are subject to the Galactic foreground absorption.
The temperature of the hot gas is allowed
to vary, but the abundance is linked among the four spectra.
We adopt the abundance standard of Grevesse and Sauval (1998)
and set a physically meaningful upper limit of 10 solar for the abundance.
The model gives a statistically acceptable fit to
all four spectra, with the overall $\chi^2/d.o.f.$ = 481.9/511.
Fit results (Table~\ref{tab:spec_fit}) suggest that the gas
temperature vary little with radius.
Interestingly, the metal abundance ($>$ 0.4 solar) is well distinguished 
from very sub-solar values that were often reported in galactic X-ray studies
(e.g., NGC~253, Strickland et al.~2002; NGC~4631, Wang et al.~2001).
We suggest that this owes to the proper modeling of the local background,
especially at energies below 0.7 keV, where the thermal continuum from the
galaxy is highly entangled with the background components.
An example of this kind has also been presented by Humphrey \& Buote (2006), who 
find near-solar iron abundances for the hot gas in most of their 
sample early-type galaxies.

We further use the PROJCT model in XSPEC to fit the spectra for a 2-D to
3-D deprojection, i.e., the fitting parameters
are measured for consecutive spherical shells.
The fit is of similar significance, with a $\chi^2/d.o.f.$ = 483.2/511.
Fit results are listed in Table~\ref{tab:spec_fit_3d}, again indicating
a quasi-isothemal hot gas with marginally super-solar abundance
in the bulge of \xs.
\begin{figure}[h] 
\centerline{
\psfig{figure=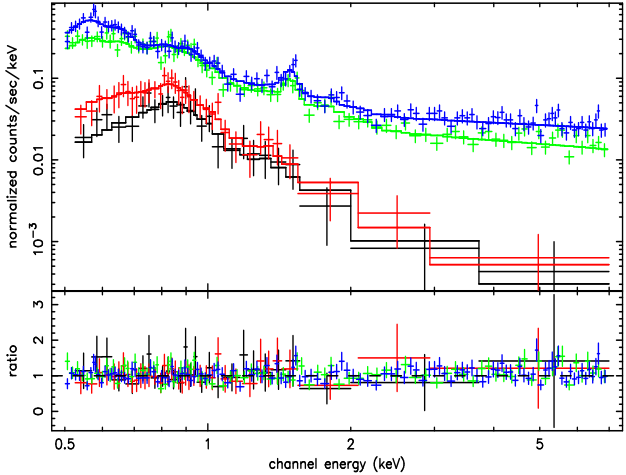,width=0.5\textwidth,angle=0, clip=}
}
\caption{Spectra of the source-subtracted emission of M~104, extracted
from concentric annuli of 30$^{\prime\prime}$-$1^\prime$ ({\sl black}) and 
1$^\prime$-$2^\prime$ ({\sl red}) from the ACIS-S, and 2$^\prime$-$4^\prime$ ({\sl green}) 
and 4$^\prime$-$6^\prime$ ({\sl Blue}) from the PN.  The two ACIS-S spectra are ``stowed
background''-subtracted. The best-fit 3-D model (see text)
is also shown.}
\label{fig:spectra_all}
\end{figure}

The fitted amount of the PL component in individual spectrum is verified by estimating 
the contribution of unresolved galactic sources. Wang (2004) obtained the LF for the
detected galactic sources, mostly LMXBs. Assuming that this LF is also valid for sources
below the source detection limit and varies little among the regions of our spectral interest, 
the contribution of unresolved sources can be taken as the integrated flux from
the LF, weighted by the amount of K-band light within individual annulus. We note that
the integrated flux of unresolved sources in the PN is $\sim$4 times higher than that
in the ACIS-S, due to the higher source detection limit in the PN. The fitted PL 
fluxes (Table~\ref{tab:spec_fit_3d}) are consistent with the above estimation 
to within 10\% (25\%) for the ACIS-S (PN) spectra. The fluxes are also consistent with
the X-ray-to-K-band intensity ratio obtained from the spatial analysis 
(Table~\ref{tab:rbp}), given a count rate to flux conversion factor predicted by the PL.

Assuming a filling factor of unity, the mean densities of hot gas 
are $\sim$6.6, 2.8, 1.2 and 0.79 $\times10^{-3}{\rm~cm^{-3}}$ in the four consecutive shells with increasing 
radii, derived from the 3-D spectral analysis (Table~\ref{tab:spec_fit_3d}).
These are shown versus radius in Fig.~\ref{fig:gp}. 
Also shown is the density profile inferred from the best-fit deVaucouleur's law
to the radial intensity distributions (\S~\ref{subsubsec:rbp}; Young 1976), with the assumption that the 
temperature of gas is constant along with radius. This profile fairly matches
the spectral measurement, indicating consistency between
our spatial and spectral analyses. 

As shown in \S~\ref{subsubsec:center}, deviations from the assumed
axisymmetry are present in the diffuse emission, especially in the inner region. Nevertheless, even in
the innermost annulus, the deviations would only introduce an uncertainty of 30\% in the average intensity,
or $\sim$15\% in the measured density. Therefore, the presence of the moderate
deviations does not qualitatively affect the determination of the radial structure of hot gas
and its implications as we discuss below.   

The total mass of hot gas contained in the shells is $\sim$4.6 $\times10^8{\rm~M_{\odot}~yr^{-1}}$,
and the intrinsic 0.2-2 keV luminosity from our spectral extraction region
is $\sim 3.1 \times10^{39}{\rm~ergs~s^{-1}}$. 
We note that these values can be approximated as the total mass and luminosity
of hot gas in \xs, given the steep density distribution (\S~\ref{subsubsec:rbp}).
For example, the luminosity within the central 6$^\prime$ is about 75\% of the total
for a de Vaucouleur's distribution with a half-light radius of 2\farcm5.

\begin{figure}[h] 
\centerline{
\psfig{figure=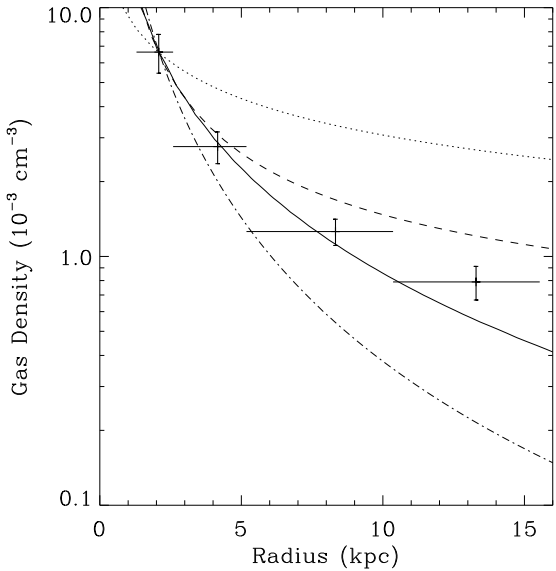,width=0.5\textwidth,angle=0, clip=}
}
\caption{
{\sl Crosses}: the measured density of hot gas versus radius; {\sl Curves}: the best-fit density profile
to the radial surface intensity distribution (solid), the predicted
density profiles of an adiabatic (dotted)  
or isothermal (dashed) gaseous corona in hydrostatic equilibrium and an 1-D steady galactic wind (dot-dashed). 
The profiles are assumed to be equal to the measurement at the first bin. See \S~\ref{sec:gas} for details.
}
\label{fig:gp}
\end{figure}

\section{Discussion} \label{sec:gas}

\subsection{The thermal structure of hot gas} \label{subsec:property}

We further compare the measured density profile with that predicted from variant
thermal structures that may be assumed for the hot gas. 
One commonly assumed case is that the gas is in hydrostatic equilibrium, i.e., the density profile
is simply determined by the gravitational potential and the equation of state for
the gas. A second case is that the gas is in the form of a large-scale outflow,
i.e., a galactic wind (e.g., Mathews \& Baker 1971; Bregman 1980; White \& Chevalier 1983),
in which the physical structure of the gas is regulated by the energy and mass input from
the stellar content.

We first characterize the 1-D distribution of the gravitational mass for the galactic bulge and halo.
The stellar distribution in projection 
is assumed to follow the de Vaucouleur's law
with a half-light radius of 1$^\prime$ ($\sim 2.6 {\rm~kpc}$; Jarrett et al.~2003).
Given the K-band magnitude of \xs,
the total stellar mass is estimated to be
1.6$\times10^{11}{\rm~M_{\odot}}$, according to the K-band mass-to-light
relation of Bell \& de Jong (2001).
However, Bridges et al.~(1997) found a kinematic mass about three times of this value
within a projected radius of $\sim$ 15 kpc.
Therefore, a second gravitational component, i.e., dark matter,
is considered. We assume that the distribution of dark matter 
follows the NFW profile (Navarro, Frenk \& White 1996) with a scale radius of 20 kpc.
This chosen scale radius is rather arbitrary but typical for a galactic dark matter halo.
The mass density of the dark matter is such that
the total mass (stars and dark matter) within a projected radius of 15 kpc be 5$\times10^{11}{\rm~M_{\odot}}$.  

For the hydrostatic equilibrium case, the density profile is calculated
for two plausible states of the gas: adiabatic and an isothermal. For the 
galactic wind case, we have developed a simple 1-D steady dynamical model (Li \& Wang in preparation). Given the 
above distributions of stars and dark matter, the density
profile of a galactic wind is determined by the rates of total energy and mass inputs to the gas which can be 
estimated from empirical measurements (see \S~\ref{subsec:feedback}). 
The measured and modeled density profiles are together shown in Fig.~\ref{fig:gp},
in which the modeled ones are assumed to match the measurement at the first bin.
We note that there is a remarkable degeneracy between the measured density and metal abundance.
Since we have assumed a single abundance independent of the radius, the overall shape 
of the density profile is not affected by this density-abundance degeneracy. 

Fig.~\ref{fig:gp} shows that an adiabatic gas in hydrostatic equilibrium (dotted curve) is inconsistent with the 
measurement, whereas neither an isothermal gas in hydrostatic equilibrium (dashed curve) nor a galactic wind 
(dot-dashed curve) can be completely ruled out
given the relatively large uncertainties in the measurement. The isothermal case is favored
by the measured temperature profile with little variation, but is subject to
further considerations in which stellar feedback to the hot gas is involved (see below). A galactic wind predicts a decreasing temperature with increasing
radius (e.g., Chevalier \& Clegg 1985) and is apparently in contradiction with the temperature
measurement. However, in the galactic wind case,  
gas moves rapidly outward. Beyond certain radius, the recombination
timescales of some species of ions may become longer than the local dynamical timescale.
Recombinations of such ions occur at larger radii than the
collisional ionization equilibrium (CIE) would predict.
This process is called a ``delayed reconbination'' (e.g, Ji, Wang \& Kwan 2006).
Thermal processes from regions involving delayed reconbinations are of non-equilibrium ionization
(NEI). Usually, when using a CIE plasma emission model (e.g., APEC)
to fit an X-ray spectrum of hot gas, temperature is effectively determined 
by the position of line features that are prominent in the spectrum. Therefore, had the NEI emission from a galactic wind been observed, the measured temperature
with a CIE model would be higher than the local gas temperature of the region
being observed. Also, the emission measure
would be higher than what the density distribution of the wind predicts. 
We suggest that this might be the case in \xs. 
A quantitative study of the NEI emission from galactic winds is under investigation. 

We now turn to discuss the origin of the hot gas in specific scenarios.


\subsection{Accretion from the intergalactic medium?} {\label{subsec:accretion}}

One scenario that may favor a quasi-hydrostatic gaseous halo comes from a specific prediction of
current theories of galaxy formation and evolution: the 
gas is accreted from the intergalactic medium (IGM) and maintained in the halos of spiral galaxies. 
The IGM is supposedly heated to X-ray-emitting temperatures, 
chiefly due to accretion shocks and gravitational compression (e.g.,
Toft et al.~2002). 
Naturally, the X-ray luminosity of gas 
is a strong function of the gravitational mass of the host galaxy, 
as characterized by its circular rotation speed ($L_X \propto V_c^7$; Toft et al.~2002). 

Surprisingly, there is little direct observational evidence for the
presence of such X-ray-emitting halos even around massive spiral galaxies.
Given the high circular rotation speed of \xs\ ($\sim 370 {\rm~km~s^{-1}}$),
it is a good candidate to look for X-ray signals from an accreted gaseous halo around it.
Toft et al.~(2002, therein Fig.~3) predict a 0.2-2 keV X-ray luminosity of
$\sim 10^{41}{\rm~ergs~s^{-1}}$
for galaxies with circular rotation speeds similar to that of \xs, about 95\%
of which coming from within 20 kpc of the disk.
However, the observed 0.2-2 keV diffuse X-ray luminosity from \xs\ is only 
$\sim 3\times10^{39}{\rm~ergs~s^{-1}}$ within the central $\sim$ 15 kpc. 
Therefore, there is a remarkable discrepancy between 
the amount of observed extraplanar hot gas and that predicted by numerical simulations.

If the hot gas in \xs\ is indeed accreted from the IGM, 
what may be the cause for this discrepancy? One plausible answer is that
the existing simulations have not adequately accounted for the feedback 
from galaxies, especially the heating due to Type Ia supernovae (SNe). 
Such stellar feedback tends to provide an effective form of large-scale distributed heating
and thus reduce the cooling of gas in the galactic bulges and halos.
If the feedback is strong enough, the galaxy may even cease further accretion.



\subsection{Stellar feedback from \xs} {\label{subsec:feedback}}

Empirically, stars in a galactic bulge continuously 
deposit energy and mass 
to the interstellar medium (ISM) 
at rates of $\sim$$6\times10^{40}$$[L_B/(10^{10} L_{B,\odot})]$
${\rm~ergs~s^{-1}}$ and $\sim$$0.2$$[L_B/(10^{10} L_{B,\odot})]$
${\rm~M_\odot~yr^{-1}}$ 
(e.g., Mannucci et al. 2005, Knapp, Gunn \& Wynn-Williams 1992), respetively,
where $L_B$ is the blue luminosity of the bulge. 
Both the stellar mass loss and the Type Ia SN rates are believed to be substantially 
greater at high redshifts when the bulges are young (e.g., Ciotti et al. 1991). 
Meanwhile, if the metals contained in the stellar ejecta are uniformly mixed
with the ISM, the mean iron abundance of the ISM is expected
to be $Z_{Fe} = Z_{*,Fe}+ 7.4(M_{Fe}/0.7 {\rm M_\odot})$,
where $M_{Fe}$ is the iron mass yield per Type Ia SN (e.g., Nomoto, Thielemann \& Yokoi 1984)
and a solar iron-to-hydrogen ratio in number of 3.16$\times10^{-5}$ is 
adopted (Grevesse and Sauval 1998).

Had most of the stellar feeback been retained by the ISM in the galaxy 
since the onset of Type Ia SNe,
it is expected that the observed X-ray luminosity and mass of hot gas be
the amount inferred from the above energy and mass input rates.
In the case of \xs, $L_B = 3.8 \times10^{10} L_{B,\odot}$, corresponding to
an energy input rate of $\sim$2.4$\times10^{41}{\rm~ergs~s^{-1}}$ and 
a mass input rate of $\sim$0.8${\rm~M_\odot~yr^{-1}}$, 
or a total mass input of 8$\times10^9{\rm~M_\odot}$ over a period of 10 Gyr.
However, our measurement (\S~\ref{subsec:property}) shows
that the rate of energy released from and the mass contained in the hot gas 
of \xs\ are nearly two orders of magnitude lower than the empirical expectations.
Given the prominent Fe L-shell features in the spectra, the fitted metal abundance
should be largely weighted by the abundance of iron. Hence the fitted value is also lower
than the empirical expectation, if the iron ejected by the SNe is uniformly 
distributed into the ISM. It is worth to note that metal abundance can easily be
under-estimated in the spectral analysis of X-ray CCD data, especially with 
over-simplified models (e.g., in the case of NGC~1316 as demonstrated 
by Kim \& Fabbiano 2003). Nevertheless, the lack of metals in \xs\ is 
evident and mostly tied to the small amount of X-ray emitting gas. 
Overall, there is a ``missing stellar feedback'' problem in \xs.

In fact, this ``missing stellar feedback'' problem is often met in the 
so-called low $L_X/L_B$ early-type galaxies 
(typically Sa spirals, S0, and low mass ellipticals), 
where the X-ray luminosity, mass and metal content of the hot gas inferred
from observations represent only a small fraction of what is
expected from the stellar feedback
(Irwin, Sarazin \& Bregman 2002; O'Sullivan, Ponman \& Collins 2003). 
These discrepancies are a clear indication for Type Ia SN-driven galactic winds (e.g., Irwin et al. 2002; Wang 2005).
Globally, winds can continueously transport the bulk of stellar depositions into the IGM, leaving
only a small fraction to be revealed within the optical extent of the host galaxy. Locally,
our analysis (\S~\ref{subsec:property}) for \xs\ indeed shows that
the thermal structure of a wind is reasonably consistent with the observation, although more detailed considerations involving NEI processes in the gas are likely needed.

\subsection{Feedback from the central AGN} {\label{subsec:AGN}}
Feedback from AGNs is a potential and sometimes favorable mechanism to affect the 
accretion of the IGM and the structure of hot gas. 
This is suggested to be the case in \xs\ (Pellegrini et al.~2003), even though its AGN has only a very sub-Eddington luminosity.
AGN feedback, if present, would disturb the gas distribution in the circumnuclear region. 
For example, dips seen in the X-ray intensity distributions at certain azimuthal ranges (Figs.~\ref{fig:ha_x} 
and \ref{fig:surbaz}) might be the result of hot gas removal by the collimated ejecta from the AGN.
However, no strong evidence of such collimated ejecta is seen in the radio continumm map
of \xs~(Bajaja et al.~1988).
Furthermore, the inclusion of the AGN feedback would only increase the energy discrepancy discussed above.
Therefore, although the possibility of AGN feedback can not be ruled out, we suggest that it plays little role
in regulating the large-scale structure of hot gas in \xs.

\section{Summary}

We have conducted a systematic analysis of the \xmm\  and \chandra\ X-ray 
observations of the nearby massive Sa galaxy \xs. The main results of our 
analysis are as follows:
\begin{itemize}

\item  We have detected large-scale diffuse X-ray emission around \xs\
to an extent of $\sim$ 20 kpc from the galactic center, which is substantially
more extented than the stellar content;

\item While at large scale the distribution of the diffuse X-ray emission tends
to be smooth, intensity fluctuations are present in the inner region;

\item Our spectral analysis of the diffuse emission reveals a gas temperature of $\sim$ 0.6-0.7 keV,
with little spatial variation, while the measured gas density drops with increasing radius, 
in a way apparently different from the expected density distribution of either an isothermal gas 
in hydrostatic equilibrium or a galactic wind, assuming CIE emission;

\item We have compared our measurements with the predictions
of numerical simulations of galaxy formation and find that 
the observed 0.2-2 keV luminosity 
($\sim 3.3 \times 10^{39} {\rm~erg~s^{-1}}$) is substantially lower than the predicted value; 

\item We have further compared the mass, energy, and metal contents 
of the hot gas with the expected inputs from the stellar feedback in \xs.
Much of the feedback is found to be missing, 
as is the case in some other X-ray faint early-type galaxies. 
A logical solution for this missing stellar feedback problem is the presence of a galactic wind, 
driven primarily by Type Ia SNe.
\end{itemize}


\acknowledgements
We thank B. Farley for assistance in the initial data analysis and L. Ji for helpful comments. 
This work is supported by the NASA grants NNG05GC69G and G06-7069X.


\begin{deluxetable}{lr}
\tablecaption{Basic Information of M~104}
\tablewidth{0pt}
\tablehead{
\colhead{Parameter} &
\colhead{M~104}}
\startdata
Morphology$^a$     \dotfill & SA(s)a \\
Center position$^a$ \dotfill & R.A.~$12^{\rm h} 39^{\rm m} 59\fs43$  \\
~~ (J2000) \dotfill & Dec.~$-11^\circ 37^\prime 23\farcs0$ \\
Optical size$^a$\dotfill & $8\farcm7\times3\farcm5$ \\
Inclination angle$^b$  \dotfill & 84$^\circ$\\
B-band magnitude $^a$  \dotfill &  8.98 \\
V-band magnitude $^a$ \dotfill & 8.00 \\
K-band magnitude $^a$ \dotfill & 4.96 \\
Circular speed (${\rm km~s^{-1}}$)$^c$\dotfill  & $304$\\
Distance (Mpc)$^d$ \dotfill & $8.9$ \\
\dotfill & ($1^\prime~\cor~$~2.59kpc) \\
Redshift$^a$ \dotfill & $0.00342$\\
Galactic foreground $N_{\rm \HI}$ ($10^{20}{\rm~cm^{-2}}$)$^e$\dotfill &$3.7$ \\
\enddata
\tablerefs{
 $a.~$NED;
 $b.~$Rubin et al.~(1985);
 $c.~$Wagner, Dettmar \& Bender (1989);
 $d.~$Ford et al.~(1996);
 $e.~$Dickey \& Lockman (1990).
}
\label{tab:M104}
\end{deluxetable}

\begin{deluxetable}{lrrrrrrrr}
  \tabletypesize{\footnotesize}
  \tablecaption{{\sl Chandra} Source List \label{tab:acis_source_list}}
  \tablewidth{0pt}
  \tablehead{
  \colhead{Source} &
  \colhead{CXOU Name} &
  \colhead{$\delta_x$ ($''$)} &
  \colhead{CR $({\rm~cts~ks}^{-1})$} &
  \colhead{HR} &
  \colhead{HR1} &
  \colhead{Flag} \\
  \noalign{\smallskip}
  \colhead{(1)} &
  \colhead{(2)} &
  \colhead{(3)} &
  \colhead{(4)} &
  \colhead{(5)} &
  \colhead{(6)} &
  \colhead{(7)} 
  }
  \startdata
   1 &  J123929.79-114549.9 &  1.2 &$     9.32  \pm   1.89$& --& $-0.37\pm0.17$ & B \\
   2 &  J123930.90-114605.4 &  1.5 &$     3.88  \pm   1.18$& --& --& S \\
   3 &  J123937.69-114031.0 &  0.5 &$     3.67  \pm   0.91$& --& --& B \\
   4 &  J123938.67-113851.3 &  0.6 &$     2.01  \pm   0.72$& --& --& B \\
   5 &  J123941.88-113725.6 &  0.8 &$     0.48  \pm   0.21$& --& --& S \\
   6 &  J123942.11-114229.1 &  1.9 &$     0.37  \pm   0.23$& $ 1.00\pm0.14$ & --& H \\
   7 &  J123942.42-113656.2 &  0.4 &$     1.46  \pm   0.34$& --& --& B \\
   8 &  J123943.20-113644.3 &  0.5 &$     1.07  \pm   0.28$& --& --& B \\
   9 &  J123943.27-114550.8 &  2.4 &$     1.23  \pm   0.50$& --& --& B \\
  10 &  J123943.61-114033.7 &  0.2 &$    15.29  \pm   1.92$& $-0.39\pm0.12$ & $-0.30\pm0.11$ & B \\
  11 &  J123943.71-113502.5 &  2.9 &$     1.30  \pm   0.64$& --& --& S \\
  12 &  J123944.17-113600.4 &  0.3 &$     3.74  \pm   0.54$& $-0.27\pm0.16$ & $ 0.52\pm0.17$ & B \\
  13 &  J123944.45-114115.0 &  0.7 &$     0.73  \pm   0.26$& --& --& B \\
  14 &  J123944.57-113727.7 &  0.5 &$     0.60  \pm   0.22$& --& --& B \\
  15 &  J123945.21-113849.5 &  0.1 &$    55.46  \pm   2.11$& $-0.77\pm0.03$ & $ 0.35\pm0.04$ & S \\
  16 &  J123945.26-113600.5 &  0.2 &$     6.46  \pm   0.70$& $ 0.11\pm0.14$ & $ 0.29\pm0.16$ & B \\
  17 &  J123945.62-113933.5 &  0.6 &$     1.32  \pm   0.49$& --& --& B \\
  18 &  J123946.34-114113.4 &  0.3 &$     3.50  \pm   0.83$& --& --& B \\
  19 &  J123946.35-114011.9 &  2.5 &$     0.86  \pm   0.52$& --& --& S \\
  20 &  J123947.21-113814.6 &  0.3 &$     1.44  \pm   0.34$& --& --& B \\
  21 &  J123947.27-114225.8 &  1.7 &$     0.81  \pm   0.42$& --& --& S \\
  22 &  J123947.68-114647.0 &  1.7 &$     1.75  \pm   0.75$& --& --& S \\
  23 &  J123948.41-113354.4 &  0.7 &$     1.17  \pm   0.33$& --& --& B \\
  24 &  J123948.43-114354.8 &  0.9 &$     1.12  \pm   0.48$& --& --& B \\
  25 &  J123948.57-113715.9 &  0.5 &$     0.88  \pm   0.27$& --& --& B \\
  26 &  J123948.61-113713.0 &  0.1 &$    11.73  \pm   0.94$& $-0.25\pm0.10$ & $ 0.31\pm0.10$ & B \\
  27 &  J123948.77-114008.5 &  0.3 &$     1.68  \pm   0.61$& --& --& B \\
  28 &  J123949.16-114340.0 &  0.3 &$    10.43  \pm   1.70$& $-0.36\pm0.16$ & $-0.34\pm0.14$ & B \\
  29 &  J123949.51-113845.7 &  0.7 &$     0.37  \pm   0.16$& --& --& B \\
  30 &  J123950.09-113741.7 &  0.3 &$     0.56  \pm   0.21$& --& --& B \\
  31 &  J123950.50-113914.4 &  0.1 &$     3.14  \pm   0.47$& $ 0.02\pm0.19$ & --& B \\
  32 &  J123950.55-113807.3 &  0.6 &$     0.42  \pm   0.19$& --& --& B \\
  33 &  J123950.64-114122.6 &  0.9 &$     0.67  \pm   0.40$& --& --& B \\
  34 &  J123950.68-114152.5 &  1.8 &$     0.33  \pm   0.18$& --& --& B \\
  35 &  J123950.87-114148.1 &  0.4 &$     1.28  \pm   0.46$& --& --& B \\
  36 &  J123950.94-113823.4 &  0.1 &$     6.38  \pm   0.67$& $ 0.04\pm0.13$ & $ 0.68\pm0.13$ & B \\
  37 &  J123951.08-113647.7 &  0.3 &$     1.34  \pm   0.32$& --& --& B \\
  38 &  J123951.11-113928.3 &  0.4 &$     0.48  \pm   0.18$& --& --& B \\
  39 &  J123951.11-113833.7 &  0.3 &$     1.14  \pm   0.29$& --& --& B \\
  40 &  J123951.97-113551.9 &  0.2 &$     5.14  \pm   0.66$& $-0.39\pm0.15$ & $ 0.56\pm0.14$ & B \\
  41 &  J123952.03-113710.6 &  0.4 &$     0.44  \pm   0.20$& --& --& B \\
  42 &  J123952.89-113739.1 &  0.7 &$     0.29  \pm   0.14$& --& --& H \\
  43 &  J123953.41-113709.4 &  0.3 &$     1.25  \pm   0.30$& --& --& B \\
  44 &  J123953.51-113808.1 &  1.0 &$     0.36  \pm   0.16$& --& --& B \\
  45 &  J123953.90-113537.4 &  0.6 &$     0.52  \pm   0.20$& --& --& B \\
  46 &  J123954.00-113825.1 &  0.2 &$     2.71  \pm   0.50$& --& $-0.37\pm0.16$ & B \\
  47 &  J123954.15-113713.0 &  0.3 &$     0.69  \pm   0.24$& --& --& B \\
  48 &  J123954.62-113602.7 &  0.7 &$     0.62  \pm   0.25$& --& --& B \\
  49 &  J123954.79-113730.2 &  0.5 &$     0.54  \pm   0.21$& --& --& B \\
  50 &  J123954.90-113708.6 &  0.2 &$     1.47  \pm   0.32$& --& --& B \\
  51 &  J123955.07-113737.0 &  0.2 &$     1.90  \pm   0.35$& $ 0.72\pm0.14$ & --& H \\
  52 &  J123955.43-113849.0 &  0.2 &$     2.67  \pm   0.44$& --& --& B \\
  53 &  J123955.55-113732.0 &  0.5 &$     0.67  \pm   0.23$& --& --& B \\
  54 &  J123955.81-113447.8 &  0.3 &$     3.49  \pm   0.51$& $-0.02\pm0.19$ & $ 0.33\pm0.20$ & B \\
  55 &  J123955.84-113731.5 &  0.3 &$     0.81  \pm   0.24$& --& --& B \\
  56 &  J123955.88-113742.1 &  0.4 &$     0.68  \pm   0.22$& --& --& B \\
  57 &  J123956.27-113154.7 &  0.4 &$     6.56  \pm   0.75$& $ 0.15\pm0.14$ & $ 0.28\pm0.17$ & B \\
  58 &  J123956.39-113759.1 &  0.2 &$     2.16  \pm   0.40$& --& --& B \\
  59 &  J123956.45-113724.3 &  0.2 &$     2.22  \pm   0.42$& --& --& B \\
  60 &  J123956.46-113830.5 &  0.3 &$     1.35  \pm   0.31$& --& --& B \\
  61 &  J123956.95-113658.7 &  0.3 &$     0.95  \pm   0.26$& --& --& B \\
  62 &  J123957.16-113703.9 &  0.4 &$     0.67  \pm   0.24$& --& --& B \\
  63 &  J123957.19-113633.6 &  0.4 &$     0.58  \pm   0.21$& --& --& S \\
  64 &  J123957.23-114135.6 &  0.7 &$     0.37  \pm   0.17$& --& --& H \\
  65 &  J123957.31-113629.6 &  0.7 &$     0.37  \pm   0.17$& --& --& S \\
  66 &  J123957.36-113750.3 &  0.5 &$     0.68  \pm   0.24$& --& --& B \\
  67 &  J123957.40-113719.8 &  0.1 &$     8.12  \pm   0.76$& $ 0.18\pm0.11$ & $ 0.44\pm0.14$ & B \\
  68 &  J123957.45-113752.8 &  0.3 &$     2.23  \pm   0.40$& --& $ 0.73\pm0.18$ & B \\
  69 &  J123957.46-113724.4 &  0.3 &$     1.01  \pm   0.29$& --& --& B \\
  70 &  J123957.47-113733.7 &  0.3 &$     1.07  \pm   0.30$& --& --& H \\
  71 &  J123957.60-113725.5 &  0.4 &$     0.50  \pm   0.21$& --& --& B \\
  72 &  J123957.70-113852.9 &  0.2 &$     1.92  \pm   0.37$& --& --& B \\
  73 &  J123957.74-113714.6 &  0.5 &$     0.60  \pm   0.23$& --& --& B \\
  74 &  J123957.93-114515.0 &  1.0 &$     2.13  \pm   0.68$& --& --& B \\
  75 &  J123957.94-113702.3 &  0.2 &$     2.10  \pm   0.40$& --& --& B \\
  76 &  J123958.00-113735.2 &  0.3 &$     1.82  \pm   0.37$& --& --& B \\
  77 &  J123958.08-114137.6 &  0.4 &$     0.72  \pm   0.24$& --& --& B \\
  78 &  J123958.37-113717.6 &  0.2 &$     2.08  \pm   0.41$& --& --& B \\
  79 &  J123958.50-113720.9 &  0.3 &$     0.85  \pm   0.27$& --& --& B \\
  80 &  J123958.66-113727.5 &  0.3 &$     1.28  \pm   0.31$& --& --& B \\
  81 &  J123958.68-114451.3 &  0.9 &$     2.32  \pm   0.82$& --& --& S \\
  82 &  J123958.74-113721.9 &  0.4 &$     1.21  \pm   0.31$& --& --& B \\
  83 &  J123958.75-114244.4 &  0.4 &$     1.82  \pm   0.51$& --& --& B \\
  84 &  J123958.76-113820.6 &  0.6 &$     0.63  \pm   0.22$& $-1.00\pm0.10$ & --& B \\
  85 &  J123958.78-113724.9 &  0.2 &$     3.02  \pm   0.49$& --& --& B \\
  86 &  J123958.78-113700.2 &  0.5 &$     0.73  \pm   0.24$& --& --& B \\
  87 &  J123958.91-113654.7 &  0.2 &$     1.57  \pm   0.35$& --& --& B \\
  88 &  J123958.97-113837.9 &  0.2 &$     2.69  \pm   0.43$& $-0.20\pm0.19$ & $ 0.65\pm0.18$ & B \\
  89 &  J123959.02-113723.0 &  0.3 &$     1.28  \pm   0.34$& --& --& B \\
  90 &  J123959.05-113512.2 &  0.2 &$     5.79  \pm   0.64$& $-0.04\pm0.14$ & $ 0.47\pm0.15$ & B \\
  91 &  J123959.09-113719.5 &  0.1 &$     6.08  \pm   0.70$& $-0.51\pm0.13$ & $ 0.38\pm0.13$ & B \\
  92 &  J123959.28-113828.1 &  0.2 &$     3.26  \pm   0.50$& $-0.04\pm0.19$ & --& B \\
  93 &  J123959.32-113650.0 &  0.4 &$     0.70  \pm   0.26$& --& --& S \\
  94 &  J123959.37-113846.0 &  0.5 &$     0.56  \pm   0.21$& --& --& B \\
  95 &  J123959.45-113727.0 &  0.1 &$     6.34  \pm   0.70$& $-0.25\pm0.14$ & $ 0.36\pm0.14$ & B \\
  96 &  J123959.45-113722.8 &  0.0 &$   147.72  \pm   3.16$& $ 0.16\pm0.03$ & $ 0.69\pm0.03$ & B\\
  97 &  J123959.53-114538.0 &  1.6 &$     0.46  \pm   0.27$& $ 1.00\pm0.19$ & --& H \\
  98 &  J123959.54-113721.2 &  0.1 &$    15.17  \pm   1.04$& $ 0.42\pm0.07$ & $ 0.28\pm0.12$ & B \\
  99 &  J123959.55-113701.8 &  0.1 &$     3.67  \pm   0.51$& $-0.06\pm0.17$ & $ 0.54\pm0.18$ & B \\
 100 &  J123959.69-113524.9 &  0.3 &$     3.81  \pm   0.51$& $-0.01\pm0.17$ & $ 0.54\pm0.18$ & B \\
 101 &  J123959.78-113716.1 &  0.2 &$     2.85  \pm   0.51$& --& $-0.04\pm0.19$ & B \\
 102 &  J123959.81-113700.2 &  0.2 &$     1.25  \pm   0.30$& --& --& B \\
 103 &  J123959.81-113455.4 &  0.3 &$     1.62  \pm   0.35$& --& --& B \\
 104 &  J123959.88-113723.5 &  0.3 &$     1.29  \pm   0.36$& --& --& B \\
 105 &  J123959.91-113622.9 &  0.1 &$     6.36  \pm   0.75$& $-0.16\pm0.15$ & $ 0.35\pm0.15$ & B \\
 106 &  J124000.04-113726.9 &  0.3 &$     1.37  \pm   0.35$& --& --& B \\
 107 &  J124000.06-113708.5 &  0.1 &$     5.12  \pm   0.61$& $-0.05\pm0.15$ & $ 0.54\pm0.15$ & B \\
 108 &  J124000.07-113609.4 &  1.3 &$     0.50  \pm   0.21$& --& --& B \\
 109 &  J124000.11-113226.7 &  1.1 &$     0.71  \pm   0.26$& --& --& B \\
 110 &  J124000.15-113542.0 &  0.7 &$     0.58  \pm   0.26$& --& --& S \\
 111 &  J124000.19-113722.5 &  0.2 &$     3.27  \pm   0.51$& $-0.07\pm0.19$ & --& B \\
 112 &  J124000.36-113723.2 &  0.1 &$     6.67  \pm   0.74$& $-0.14\pm0.15$ & $ 0.06\pm0.14$ & B \\
 113 &  J124000.46-113717.6 &  0.4 &$     1.06  \pm   0.32$& --& --& B \\
 114 &  J124000.61-114343.4 &  0.4 &$     5.22  \pm   1.07$& $ 0.28\pm0.19$ & --& B \\
 115 &  J124000.65-113110.9 &  1.5 &$     1.63  \pm   0.58$& --& --& B \\
 116 &  J124000.68-113712.5 &  0.3 &$     0.91  \pm   0.28$& --& --& B \\
 117 &  J124000.69-113519.5 &  0.2 &$     5.38  \pm   0.60$& $ 0.03\pm0.14$ & $ 0.59\pm0.15$ & B \\
 118 &  J124000.70-113704.2 &  0.2 &$     2.71  \pm   0.45$& --& --& B \\
 119 &  J124000.72-114602.4 &  1.0 &$     3.17  \pm   0.88$& --& --& B \\
 120 &  J124000.75-113730.3 &  0.3 &$     0.67  \pm   0.25$& --& --& B \\
 121 &  J124000.95-113653.9 &  0.1 &$    19.34  \pm   1.20$& $-0.32\pm0.08$ & $ 0.39\pm0.07$ & B \\
 122 &  J124000.96-113701.5 &  0.2 &$     1.33  \pm   0.32$& --& --& B \\
 123 &  J124000.98-113708.3 &  0.2 &$     1.17  \pm   0.31$& --& --& B \\
 124 &  J124001.08-113723.8 &  0.1 &$    34.17  \pm   1.72$& $-0.82\pm0.04$ & $ 0.02\pm0.05$ & B \\
 125 &  J124001.15-113722.0 &  0.5 &$     1.04  \pm   0.31$& --& --& B \\
 126 &  J124001.28-113701.9 &  0.3 &$     0.83  \pm   0.25$& --& --& B \\
 127 &  J124001.29-113729.6 &  0.2 &$     2.01  \pm   0.38$& --& --& B \\
 128 &  J124001.30-113720.1 &  0.4 &$     0.72  \pm   0.27$& --& --& B \\
 129 &  J124001.44-114010.4 &  0.7 &$     0.43  \pm   0.17$& --& --& B \\
 130 &  J124002.00-113940.2 &  0.1 &$     5.88  \pm   0.76$& $-1.00\pm0.07$ & $-0.65\pm0.08$ & B \\
 131 &  J124002.02-113707.0 &  0.3 &$     1.12  \pm   0.29$& --& --& B \\
 132 &  J124002.06-113846.5 &  0.5 &$     0.51  \pm   0.20$& --& --& B \\
 133 &  J124002.16-113723.6 &  0.5 &$     0.42  \pm   0.18$& --& --& H \\
 134 &  J124002.17-113754.0 &  0.2 &$     2.01  \pm   0.39$& --& --& B \\
 135 &  J124002.23-113718.4 &  0.2 &$     1.84  \pm   0.37$& --& --& B \\
 136 &  J124002.27-113801.6 &  0.4 &$     0.89  \pm   0.26$& --& --& H \\
 137 &  J124002.28-113711.7 &  0.3 &$     0.97  \pm   0.27$& --& --& B \\
 138 &  J124002.43-113852.1 &  0.4 &$     1.04  \pm   0.28$& --& --& B \\
 139 &  J124002.50-113647.4 &  0.7 &$     0.46  \pm   0.20$& --& --& B \\
 140 &  J124002.77-113716.7 &  0.2 &$     1.43  \pm   0.33$& --& --& B \\
 141 &  J124002.81-114250.6 &  0.4 &$     3.42  \pm   0.91$& --& --& B \\
 142 &  J124003.02-113807.6 &  0.6 &$     0.45  \pm   0.19$& --& --& B \\
 143 &  J124003.15-114004.7 &  0.3 &$     1.17  \pm   0.28$& --& --& B \\
 144 &  J124003.64-113739.2 &  0.2 &$     1.36  \pm   0.30$& --& --& B \\
 145 &  J124003.66-113242.0 &  0.6 &$     2.38  \pm   0.43$& --& $ 0.92\pm0.15$ & B \\
 146 &  J124003.72-113700.9 &  0.4 &$     0.83  \pm   0.26$& --& --& S \\
 147 &  J124004.54-113637.6 &  0.3 &$     1.95  \pm   0.41$& --& --& B \\
 148 &  J124004.62-113735.6 &  0.2 &$     1.41  \pm   0.31$& --& --& B \\
 149 &  J124004.67-113828.8 &  0.3 &$     0.71  \pm   0.23$& --& --& B \\
 150 &  J124004.81-113720.9 &  0.2 &$     1.56  \pm   0.37$& --& --& B \\
 151 &  J124005.01-113732.3 &  0.2 &$     1.78  \pm   0.35$& --& --& B \\
 152 &  J124005.36-113559.7 &  0.3 &$     3.54  \pm   0.49$& $ 0.09\pm0.17$ & $ 0.67\pm0.18$ & B \\
 153 &  J124005.54-113940.4 &  0.2 &$     2.10  \pm   0.41$& --& --& B \\
 154 &  J124005.62-114147.3 &  0.3 &$     6.56  \pm   1.23$& $-0.44\pm0.16$ & $-0.17\pm0.19$ & B \\
 155 &  J124005.70-113711.0 &  0.2 &$     2.79  \pm   0.46$& --& --& B \\
 156 &  J124006.18-113609.0 &  0.4 &$     0.54  \pm   0.20$& --& --& B \\
 157 &  J124006.32-113647.4 &  0.7 &$     0.49  \pm   0.20$& --& --& B \\
 158 &  J124006.96-113721.9 &  0.5 &$     0.66  \pm   0.23$& --& --& S \\
 159 &  J124007.05-113753.3 &  0.2 &$     3.97  \pm   0.54$& $ 0.21\pm0.17$ & --& B \\
 160 &  J124008.27-113446.2 &  0.9 &$     0.94  \pm   0.28$& --& --& B \\
 161 &  J124008.38-113711.3 &  0.4 &$     1.04  \pm   0.29$& --& --& B \\
 162 &  J124008.89-113817.3 &  0.4 &$     1.15  \pm   0.34$& --& --& B \\
 163 &  J124009.44-114154.1 &  0.7 &$     1.18  \pm   0.47$& --& --& B \\
 164 &  J124009.56-113645.8 &  0.3 &$     3.52  \pm   0.51$& $-0.02\pm0.19$ & $ 0.33\pm0.20$ & B \\
 165 &  J124010.21-113159.8 &  1.6 &$     2.34  \pm   0.98$& --& --& S \\
 166 &  J124010.44-113638.7 &  0.2 &$     6.29  \pm   0.67$& $-0.15\pm0.14$ & $ 0.54\pm0.13$ & B \\
 167 &  J124010.63-113741.2 &  0.6 &$     0.54  \pm   0.20$& --& --& B \\
 168 &  J124010.83-113258.1 &  0.5 &$     3.44  \pm   0.52$& $ 0.07\pm0.18$ & $ 0.96\pm0.14$ & B \\
 169 &  J124011.46-113444.2 &  0.5 &$     1.88  \pm   0.37$& --& --& B \\
 170 &  J124012.34-114005.3 &  0.5 &$     1.35  \pm   0.32$& --& --& B \\
 171 &  J124012.70-113630.2 &  0.5 &$     0.77  \pm   0.25$& --& --& B \\
 172 &  J124012.88-113903.1 &  0.6 &$     1.05  \pm   0.28$& --& --& B \\
 173 &  J124013.72-113752.7 &  0.5 &$     1.42  \pm   0.35$& --& --& B \\
 174 &  J124015.18-113307.5 &  1.1 &$     4.53  \pm   0.89$& --& --& B \\
 175 &  J124015.45-113736.4 &  0.8 &$     0.76  \pm   0.25$& --& --& B \\
\enddata
\tablecomments{The full table is available electronically.
The definition of the bands:
0.3--0.7 (S1), 0.7--1.5 (S2), 1.5--3 (H1), and 3--7~keV (H2). 
In addition, S=S1+S2, H=H1+H2, and B=S+H.
 Column (1): Generic source number. (2): 
{\sl Chandra} X-ray Observatory (unregistered) source name, following the
{\sl Chandra} naming convention and the IAU Recommendation for Nomenclature
(e.g., http://cdsweb.u-strasbg.fr/iau-spec.html). (3): Position 
uncertainty (1$\sigma$) (1$\sigma$) calculated from the maximum likelihood centroiding.  (4): On-axis source broad-band count rate --- the sum of the 
exposure-corrected count rates in the four
bands. (5-6): The hardness ratios defined as 
${\rm HR}=({\rm H-S2})/({\rm H+S2})$, and ${\rm HR1}=({\rm S2-S1})/{\rm S}$, 
listed only for values with uncertainties less than 0.2.
(7): The label ``B'', ``S'', or ``H'' mark the band in 
which a source is detected with the most accurate position that is adopted in
Column (3). 
}
  \end{deluxetable}
  \vfill
\eject

\begin{deluxetable}{lrrrrrrrr}
  \tabletypesize{\footnotesize}
  \tablecaption{{\sl XMM-Newton} Source List \label{tab:pn_source_list}}
  \tablewidth{0pt}
  \tablehead{
  \colhead{Source} &
  \colhead{XMMU Name} &
  \colhead{$\delta_x$ ($''$)} &
  \colhead{CR $({\rm~cts~ks}^{-1})$} &
  \colhead{HR} &
  \colhead{HR1} &
  \colhead{Flag} \\
  \noalign{\smallskip}
  \colhead{(1)} &
  \colhead{(2)} &
  \colhead{(3)} &
  \colhead{(4)} &
  \colhead{(5)} &
  \colhead{(6)} &
  \colhead{(7)} 
  }
  \startdata
   1 &  J123854.66-113734.9 &  7.3 &$    13.88  \pm   4.02$& --& --& B \\
   2 &  J123900.58-113459.2 &  4.8 &$    73.65  \pm   7.50$& $-0.28\pm0.17$ & $ 0.07\pm0.11$ & B \\
   3 &  J123910.21-114006.6 &  4.8 &$    15.24  \pm   3.41$& --& --& B \\
   4 &  J123911.34-114558.8 &  7.1 &$    10.91  \pm   3.58$& --& --& S \\
   5 &  J123913.14-113323.2 &  4.3 &$    17.28  \pm   3.79$& --& --& S \\
   6 &  J123913.46-113143.2 &  4.6 &$    19.66  \pm   3.76$& --& $-0.49\pm0.20$ & B \\
   7 &  J123913.96-112712.6 &  5.1 &$    25.39  \pm   4.80$& --& $ 0.30\pm0.19$ & S \\
   8 &  J123922.15-114718.2 &  7.4 &$     7.59  \pm   3.26$& --& --& S \\
   9 &  J123922.56-114452.9 &  4.0 &$    48.77  \pm   5.63$& $-0.22\pm0.20$ & $-0.09\pm0.13$ & B \\
  10 &  J123926.90-114007.8 &  2.7 &$    20.29  \pm   3.11$& --& $-0.48\pm0.13$ & S \\
  11 &  J123929.68-114552.7 &  4.1 &$    25.02  \pm   3.92$& --& $ 0.19\pm0.17$ & B \\
  12 &  J123932.69-113943.0 &  3.7 &$     8.88  \pm   2.29$& --& --& B \\
  13 &  J123937.90-114029.1 &  3.6 &$     6.61  \pm   1.75$& --& --& B \\
  14 &  J123938.99-113725.1 &  2.5 &$     8.56  \pm   1.81$& --& --& B \\
  15 &  J123940.60-113256.5 &  3.0 &$     6.94  \pm   1.72$& --& --& B \\
  16 &  J123941.35-112828.8 &  2.2 &$    36.71  \pm   4.11$& $-0.56\pm0.18$ & $ 0.06\pm0.12$ & B \\
  17 &  J123942.33-113653.5 &  3.4 &$     4.87  \pm   1.44$& --& --& B \\
  18 &  J123945.24-113848.1 &  1.0 &$    58.11  \pm   4.34$& $-0.86\pm0.14$ & $-0.48\pm0.07$ & S \\
  19 &  J123945.29-113600.7 &  1.5 &$    18.77  \pm   2.34$& $ 0.10\pm0.18$ & $-0.10\pm0.15$ & B \\
  20 &  J123945.73-112835.3 &  3.1 &$    13.53  \pm   2.72$& --& $-0.52\pm0.17$ & S \\
  21 &  J123948.60-113714.1 &  1.6 &$    22.64  \pm   2.88$& $-0.21\pm0.16$ & $ 0.51\pm0.14$ & B \\
  22 &  J123949.17-114337.7 &  2.6 &$    19.01  \pm   2.76$& --& $ 0.22\pm0.18$ & B \\
  23 &  J123950.68-113912.7 &  3.0 &$     6.10  \pm   1.52$& --& --& B \\
  24 &  J123950.86-114140.9 &  3.1 &$     5.69  \pm   1.52$& --& --& B \\
  25 &  J123950.96-113823.6 &  1.8 &$    13.23  \pm   2.04$& --& $-0.21\pm0.16$ & B \\
  26 &  J123954.06-113826.5 &  3.0 &$     3.56  \pm   1.60$& --& --& S \\
  27 &  J123955.47-113847.0 &  2.4 &$     9.44  \pm   1.75$& --& --& B \\
  28 &  J123956.28-113153.3 &  2.3 &$     9.03  \pm   1.77$& --& --& B \\
  29 &  J123957.38-114134.4 &  2.0 &$    16.89  \pm   2.29$& $ 0.39\pm0.13$ & --& B \\
  30 &  J123958.10-113124.1 &  3.6 &$     3.81  \pm   1.37$& --& --& S \\
  31 &  J123959.14-112531.0 &  4.9 &$     9.63  \pm   2.62$& --& --& B \\
  32 &  J123959.24-113514.1 &  1.2 &$    20.96  \pm   2.34$& $-0.30\pm0.16$ & $ 0.26\pm0.12$ & B \\
  33 &  J123959.47-113722.4 &  0.3 &$   614.97  \pm  11.64$& $-0.16\pm0.03$ & $ 0.22\pm0.02$ & B \\
  34 &  J123959.94-113621.9 &  1.9 &$    10.61  \pm   2.09$& --& --& B \\
  35 &  J124000.12-113522.5 &  2.3 &$    19.22  \pm   2.25$& $-0.16\pm0.19$ & $-0.13\pm0.13$ & H \\
  36 &  J124000.73-114342.6 &  2.9 &$    10.13  \pm   1.97$& --& --& B \\
  37 &  J124001.93-113938.9 &  2.1 &$     7.69  \pm   1.58$& --& $-0.91\pm0.12$ & S \\
  38 &  J124002.09-113754.0 &  1.9 &$    11.13  \pm   2.35$& --& --& B \\
  39 &  J124002.52-112622.2 &  4.3 &$     9.40  \pm   2.45$& --& --& S \\
  40 &  J124002.73-114247.2 &  2.8 &$     7.95  \pm   1.70$& --& --& S \\
  41 &  J124003.71-113240.6 &  2.6 &$     6.29  \pm   1.48$& --& --& B \\
  42 &  J124004.85-113733.3 &  1.8 &$    12.55  \pm   2.08$& $ 0.18\pm0.19$ & --& B \\
  43 &  J124005.26-112712.1 &  3.0 &$    18.68  \pm   3.04$& --& $ 0.53\pm0.17$ & B \\
  44 &  J124005.40-113500.0 &  2.1 &$     8.57  \pm   1.61$& --& --& B \\
  45 &  J124005.75-113710.6 &  2.5 &$     7.32  \pm   1.63$& --& --& B \\
  46 &  J124006.88-113749.9 &  2.7 &$     5.96  \pm   1.47$& --& --& B \\
  47 &  J124007.05-112756.9 &  3.0 &$    12.36  \pm   2.54$& --& $-0.64\pm0.16$ & S \\
  48 &  J124010.27-113639.0 &  1.3 &$    20.46  \pm   2.32$& $-0.00\pm0.17$ & $-0.10\pm0.14$ & B \\
  49 &  J124010.56-114729.8 &  5.1 &$     8.34  \pm   2.35$& --& --& B \\
  50 &  J124010.70-113256.8 &  2.0 &$    12.35  \pm   1.97$& --& --& B \\
  51 &  J124010.85-113000.0 &  2.9 &$     8.22  \pm   1.91$& --& --& B \\
  52 &  J124012.73-113903.9 &  3.3 &$     4.85  \pm   1.33$& --& --& B \\
  53 &  J124015.21-113304.2 &  1.9 &$    14.88  \pm   2.12$& $-0.16\pm0.20$ & $ 0.22\pm0.17$ & B \\
  54 &  J124016.84-114428.3 &  5.0 &$     7.53  \pm   1.92$& --& --& H \\
  55 &  J124024.01-113852.1 &  4.3 &$     3.71  \pm   1.45$& --& --& S \\
  56 &  J124024.47-114818.8 &  5.8 &$    11.04  \pm   3.13$& --& --& B \\
  57 &  J124025.74-114208.2 &  3.9 &$     7.39  \pm   1.81$& --& --& B \\
  58 &  J124027.09-114701.7 &  4.8 &$    20.44  \pm   3.49$& --& $-0.47\pm0.16$ & B \\
  59 &  J124029.34-113643.0 &  3.6 &$     8.67  \pm   1.91$& --& --& B \\
  60 &  J124031.31-113156.7 &  2.8 &$    17.65  \pm   2.76$& --& $ 0.53\pm0.17$ & B \\
  61 &  J124045.40-113918.1 &  5.4 &$     6.69  \pm   2.22$& --& --& S \\
  62 &  J124048.10-113703.7 &  4.2 &$    22.22  \pm   3.40$& --& $ 0.05\pm0.17$ & B 
\enddata
\tablecomments{The full table is available electronically. The definition of the bands:
0.5--1 (S1), 1--2 (S2), 2--4.5 (H1), and 4.5--7.5~keV (H2). 
In addition, S=S1+S2, H=H1+H2, and B=S+H.
 Column (1): Generic source number. (2): 
{\sl XMM-Newton} X-ray Observatory (unregistered) source name, following the
{\sl XMM-Newton} naming convention and the IAU Recommendation for Nomenclature
(http://cdsweb.u-strasbg.fr/iau-spec.html). (3): Position 
uncertainty (1$\sigma$) calculated from the maximum likelihood centroiding.  (4): On-axis source broad-band count rate --- the sum of the 
exposure-corrected count rates in the four
bands. (5-6): The hardness ratios defined as 
${\rm HR}=({\rm H-S2})/({\rm H+S2})$, and ${\rm HR1}=({\rm S2-S1})/{\rm S}$, 
listed only for values with uncertainties less than 0.2.
(7): The label ``B'', ``S'', or ``H'' mark the band in 
which a source is detected with the most accurate position that is adopted in
Column (3). 
}
  \end{deluxetable}
  \vfill

\begin{deluxetable}{lccrrrrr}
  \tabletypesize{\footnotesize}
  \tablecaption{Source Identification\label{tab:sou_id}}
  \tablewidth{0pt}
  \tablehead{
  \colhead{Source} &
  \colhead{ID Name} &
  \colhead{$\delta_{x,o}$ ($''$)} &
  \colhead{Magnitude} \\
  \noalign{\smallskip}
  }
  \startdata   

XA-4 & J123938.76-113851.2 &  1.2 $\pm$  0.6& J=11.4 H=11.1 K=11.0\\
XA-15 & J123945.21-113849.5 &  0.1 $\pm$  0.1& J= 8.9 H= 8.5 K= 8.4\\
XA-16 & J123945.26-113600.8 &  0.3 $\pm$  0.2& J=16.9 H=16.1 K=15.5\\
XA-57 & J123956.22-113154.7 &  0.9 $\pm$  0.4& J=16.9 H=16.1 K=15.5\\
XA-143 & J124003.13-114004.2 &  0.5 $\pm$  0.3& J=15.6 H=14.9 K=14.7\\
XA-168 & J124010.83-113258.5 &  0.3 $\pm$  0.5& J=16.0 H=15.2 K=14.9\\
XP-10 & J123926.92-114008.6 &  2.1 $\pm$  2.7& J= 9.0 H= 8.5 K= 8.4\\
XP-16 & J123941.33-112829.6 &  1.5 $\pm$  2.2& J=16.9 H=16.3 K=15.3
\enddata
\end{deluxetable}

\begin{deluxetable}{lrrrrrr}
\tabletypesize{\footnotesize}
\tablecaption{Spectral fits of discrete sources}
\tablewidth{0pt}
\tablehead{
\colhead{Source} & 
\colhead{N$_H$} &
\colhead{Photon index$^a$} &
\colhead{Temperature$^b$} &
\colhead{$\chi^2/d.o.f.$} &
\colhead{$L_X$ (0.3-7 keV)} \\
\colhead{} & 
\colhead{$10^{20}~cm^{-2}$} &
\colhead{} &
\colhead{keV} &
\colhead{} &
\colhead{$10^{39}{\rm~ergs~s^{-1}}$}

}
\startdata
XA-26         & $8.0^{+11.2}_{-8.0}$ & $2.13^{+0.16}_{-0.46}$ & - & 6.3/7 & 0.7 \\
XA-121        & $8.2^{+12.5}_{-8.2}$ & $1.87^{+0.45}_{-0.44}$ & - & 14.9/12 & 1.1 \\
XA-124        & $3.4^{+4.9}_{-3.4}$  & - & $0.19^{+0.01}_{-0.03}$ & 21.1/21 & 0.9 \\
Accumulated   & $10.7^{+2.2}_{-2.1}$ & $1.51^{+0.10}_{-0.09}$ & - & 115.2/144 &  26 
\enddata
\tablecomments{$^a$~For a power-law model. $^b$~For a black-body emission model.}
\label{tab:sou_spec}
\end{deluxetable}


\begin{deluxetable}{lrrrrrr}
\tabletypesize{\footnotesize}
\tablecaption{Fits to the radial surface brightness profiles$^a$}
\tablewidth{0pt}
\tablehead{
\colhead{Parameter} & 
\colhead{PN} &
\colhead{PN} &
\colhead{PN} &
\colhead{ACIS-S} &
\colhead{ACIS-S} &
\colhead{ACIS-S} \\
\colhead{}         &
\colhead{0.5-1 keV} &
\colhead{1-2 keV} &
\colhead{2-7.5 keV} &
\colhead{0.3-0.7 keV} &
\colhead{0.7-1.5 keV} &
\colhead{1.5-7 keV} 
}
\startdata
$\chi^2/d.o.f.$\dotfill & 108.0/93 & 46.7/69  & 12.5/22 & 32.2/43 & 56.1/62  & 6.2/9 \\
$I_{g}$ (${\rm~cts~s^{-1}~arcmin^{-2}}$)\dotfill & 4.9$^{+3.3}_{-2.6}$ & 1.1$^{+0.7}_{-0.5}$ & - & 1.3$^{+0.8}_{-0.5}$ & 2.0$^{+1.3}_{-0.9}$ & -\\
$r_{e}$ (arcmin)\dotfill & 2.6$^{+1.4}_{-0.9}$ & same & - & same & same & -\\
$^b$$I_s$ ($10^{-4} {\rm~cts~s^{-1}~arcmin^{-2}/[MJy~sr^{-1}]}$)\dotfill & 13.2  & 20.1  & 18.7$^{+4.7}_{-4.7}$  & 1.0  & 2.9 & 2.8$^{+0.5}_{-0.5}$ \\
$I_b$ ($10^{-4} {\rm~cts~s^{-1}~arcmin^{-2}}$)\dotfill & 25.6$^{+1.0}_{-0.7}$ & 7.8$^{+0.9}_{-0.9}$ & 7.5$^{+1.5}_{-1.5}$ & 11.8$^{+2.3}_{-2.3}$ & 6.7$^{+1.9}_{-1.8}$ & 2.8$^{+1.3}_{-1.2}$ 
\enddata
\tablecomments{$^a$~The profiles are
fitted by a normalized K-band profile plus a local constant background,
$I_X$(R)= $I_s$$I_K$(R) + $I_b$, for the PN 2-7.5 keV and ACIS-S 1.5-7 keV bands,
or with an additional de Vaucouleur's law, $I_X$(R)= $I_s$$I_K$(R) 
+ $I_g~e^{-7.67(R/r_e)^{1/4}}$ + $I_b$, for the softer bands. 
$^b$~The normalization factors for different bands are
related via the best-fit spectral model to the accumulated source spectrum (Table~\ref{tab:sou_spec}).}
\label{tab:rbp}
\end{deluxetable}

\begin{deluxetable}{lrrr}
\tabletypesize{\footnotesize}
\tablecaption{Fits to the vertical surface intensity profiles$^a$}
\tablewidth{0pt}
\tablehead{
\colhead{Parameter} & 
\colhead{ACIS-S} &
\colhead{ACIS-S} &
\colhead{ACIS-S} \\
\colhead{}         &
\colhead{0.3-0.7 keV} &
\colhead{0.7-1.5 keV} &
\colhead{1.5-7 keV} 
}
\startdata
$\chi^2/d.o.f.$\dotfill & 47.2/28 & 70.4/45  & 10.5/8 \\
$I_{g}$ ($10^{-4} {\rm~cts~s^{-1}~arcmin^{-2}}$)\dotfill & 160$^{+30}_{-30}$ & 110$^{+14}_{-14}$ & - \\
$^b$$z_{0}$ (arcmin)\dotfill & 0.22$^{+0.05}_{-0.05}$, 0.31$^{+0.06}_{-0.06}$ & 0.64$^{+0.08}_{-0.08}$, 0.72$^{+0.09}_{-0.09}$ & -\\
$^c$$I_s$ ($10^{-4} {\rm~cts~s^{-1}~arcmin^{-2}/[MJy~sr^{-1}]}$)\dotfill & 1.0  & 2.9 & 2.8 \\
$^c$$I_b$ ($10^{-4} {\rm~cts~s^{-1}~arcmin^{-2}}$)\dotfill & 11.8 & 6.7 & 2.8 
\enddata
\tablecomments{$^a$The 1.5-7 keV profile is
fitted by a normalized K-band profile plus a local constant background: 
$I_X$(z)= $I_s$$I_K$(z) + $I_b$. For the softer bands,
an additional exponential law is applied: $I_X$(z)= $I_s$$I_K$(z) 
+ $I_g~e^{-|z|/z_0}$ + $I_b$. $^b$The first and the second
values are for the south and north sides, respectively. $^c$ Same normalization factors
and local background rates are applied as for the radial profiles (Table~\ref{tab:rbp}).}
\label{tab:vbp}
\end{deluxetable}

\begin{deluxetable}{lcccc}
\tablecaption{2-D Fits to the spectra of source-subtracted emission$^a$}
\tablewidth{0pt}
\tablehead{
\colhead{Parameter} & 
\colhead{$30^{\prime\prime}-1^\prime$} &
\colhead{$1^\prime-2^\prime$} &
\colhead{$2^\prime-4^\prime$} &
\colhead{$4^\prime-6^\prime$}
}
\startdata
Temperature (keV)\dotfill & 0.62$^{+0.09}_{-0.09}$ & 0.59$^{+0.07}_{-0.10}$  & 0.63$^{+0.07}_{-0.06}$ & 0.78$^{+0.13}_{-0.11}$\\
Abundance (solar)\dotfill & 1.4 ($>$ 0.4) & same & same & same \\ 
Photon index \dotfill & 1.51$^b$ & same & same & same \\
Normalization (APEC; 10$^{-5}$)\dotfill & 2.3$^{+0.5}_{-0.5}$ & 3.6$^{+0.7}_{-0.7}$  & 3.8$^{+0.6}_{-0.6}$ & 2.7$^{+0.7}_{-0.8}$\\
Normalization (PL; 10$^{-5}$) \dotfill & 1.1$^{+0.8}_{-0.8}$  & 1.2$^{+1.0}_{-0.9}$ & 4.0$^{+0.8}_{-0.8}$ & 2.1$^{+1.2}_{-1.1}$ \\
$f_{0.2-2 {\rm~keV}}$ (APEC; 10$^{-14} {\rm~ergs~cm^{-2}~s^{-1}}$) & 6.2 & 9.3 & 9.9 & 7.2 \\
$f_{0.3-7 {\rm~keV}}$ (PL; 10$^{-14} {\rm~ergs~cm^{-2}~s^{-1}}$) & 7.2 & 8.0& 26.5  & 14.1
\enddata
\tablecomments{The spectra extracted from four consecutive annuli are fitted by a combined model of APEC+power-law (PL) with
the Galactic foreground absorption.}
\label{tab:spec_fit}
\end{deluxetable}

\begin{deluxetable}{lcccc}
\tablecaption{3-D Fits to the spectra of source-subtracted emission$^a$}
\tablewidth{0pt}
\tablehead{
\colhead{Parameter} & 
\colhead{$30^{\prime\prime}-1^\prime$} &
\colhead{$1^\prime-2^\prime$} &
\colhead{$2^\prime-4^\prime$} &
\colhead{$4^\prime-6^\prime$}
}
\startdata
Temperature (keV)\dotfill & 0.64$^{+0.14}_{-0.20}$ & 0.57$^{+0.14}_{-0.16}$  & 0.58$^{+0.10}_{-0.25}$ & 0.75$^{+0.10}_{-0.11}$\\
Abundance (solar)\dotfill & 1.7 ($>$ 0.4) & same & same & same \\ 
Photon index \dotfill & 1.51$^b$ & same & same & same \\
Normalization (APEC; 10$^{-5}$)\dotfill & 2.0$^{+0.7}_{-0.7}$ & 2.7$^{+0.8}_{-0.8}$  & 4.6$^{+1.1}_{-1.1}$ & 4.9$^{+1.5}_{-1.5}$\\
Normalization (PL; 10$^{-5}$) \dotfill & 1.1$^{+0.8}_{-0.8}$   & 1.2$^{+1.0}_{-0.9}$  & 4.0$^{+0.8}_{-0.9}$ & 2.2$^{+1.2}_{-1.1}$ \\
$f_{0.2-2 {\rm~keV}}$ (APEC; 10$^{-14} {\rm~ergs~cm^{-2}~s^{-1}}$) & 6.1 & 9.3 & 10.0 & 7.0 \\
$f_{0.3-7 {\rm~keV}}$ (PL; 10$^{-14} {\rm~ergs~cm^{-2}~s^{-1}}$) & 7.2 & 8.0 & 26.2 & 14.6 
\enddata
\tablecomments{The spectra extracted from four consecutive annuli are fitted by a combined model of PROJCT(APEC)+power-law (PL) with
the Galactic foreground absorption, where the emission is deprojected and the parameters
are measured for consecutive shells.}
\label{tab:spec_fit_3d}
\end{deluxetable}

\end{document}